\documentclass[12pt]{article}

\usepackage[a4paper,margin=1in]{geometry}
\usepackage{amsmath,amssymb}
\usepackage{booktabs}
\usepackage{graphicx}
\usepackage{float}
\usepackage{longtable}
\usepackage{array}
\usepackage{caption}
\usepackage{subcaption}
\usepackage[colorlinks=true,linkcolor=blue,citecolor=blue,urlcolor=blue]{hyperref}
\graphicspath{{figures/}}

\newcommand{\pct}[1]{#1\%}
\newcommand{\code}[1]{\texttt{\detokenize{#1}}}
\newcommand{\pathcode}[1]{\begingroup\scriptsize\url{#1}\endgroup}

\title{\bfseries Relief-Gated Relative Rotation for QQQ--DIA Allocation\\
\large Globally Screened Relative States, Fixed Position Mapping, Incremental Interaction Admission, and Walk-Forward Validation}
\author{Zheli Xiong\\
\normalsize Corresponding author: Zheli Xiong (zlxiong@mail.ustc.edu.cn)}
\date{}

\begin{document}
\maketitle

\begin{abstract}
This paper studies \emph{Relief-Gated Relative Rotation} (RGRR), a two-ETF rule that allocates between QQQ and DIA by mapping screened relative and macro states into a continuous QQQ weight. RGRR is economic rather than mechanical: it rotates between a growth-heavy sleeve and a Dow/value-heavy sleeve only when QQQ--DIA relative states are confirmed by rate, volatility, credit, or broad-market relief conditions. The design follows the empirical discipline of prior growth--defensive timing and cash-overlay studies \cite{xiong2026growthdefensive,xiong2026cashoverlay}. Candidate main effects and interactions are globally screened with horizon-specific HAC regressions and correlation de-duplication, then held fixed during walk-forward validation. Rolling OOS re-selects only signal-family lambdas, not the signal universe or the position mapping. The final stack contains one main effect, nine second-order interactions, and two third-order interactions. Third-order terms must also improve rolling OOS Sharpe versus the main+second-order base and survive economic-family de-duplication. The final mapping uses max tilt 0.50, \code{tau_weight}=0.75, \code{eta}=0.05, and a 10 bps one-way turnover cost.

Across the 2018, 2020, and 2022 OOS starts, RGRR improves Sharpe versus 100\% QQQ and 50/50 QQQ--DIA in every tested interval. It improves CAGR versus 50/50 in every interval, but beats 100\% QQQ on CAGR only in the 2022 OOS window. In 2018 OOS, RGRR earns 18.33\% CAGR and 0.94 Sharpe, versus 20.50\% and 0.89 for QQQ and 16.69\% and 0.86 for 50/50. In 2022 OOS, it earns 15.19\% CAGR and 0.87 Sharpe, versus 14.65\% and 0.70 for QQQ. The evidence supports RGRR as a risk-adjusted relative allocation rule, not a pure return-dominance rule. Its main practical weakness is high turnover, ranging from 354\% to 506\% annualized.
\end{abstract}


\noindent\textbf{Code and replication package:}
\begin{quote}
\footnotesize
\href{https://github.com/shaun19920309/Relief-Gated-Relative-Rotation-for-QQQ-DIA-Allocation}{\texttt{https://github.com/shaun19920309/}}\\
\href{https://github.com/shaun19920309/Relief-Gated-Relative-Rotation-for-QQQ-DIA-Allocation}{\texttt{Relief-Gated-Relative-Rotation-for-QQQ-DIA-Allocation}}
\end{quote}

\section{Introduction}

This paper asks whether a transparent statistical rule can improve the risk-adjusted allocation between QQQ and DIA. The economic premise is narrower than a broad equity-timing model. Both assets are equity ETFs. The strategy does not move to cash, bonds, or leverage. It only changes the QQQ weight and assigns the residual weight to DIA.

The motivating observation is familiar: QQQ can lead DIA strongly during technology-led expansions, but after a large QQQ--DIA relative advance, there are windows in which QQQ stalls and DIA catches up. The central empirical question is whether this intuition is stable enough to become a systematic allocation rule. A direct ``QQQ has led and is now flat'' rule is too narrow and did not survive the research process as a standalone mechanism. The final evidence instead points to a more conditional pattern: relative reversal states matter most when combined with rate relief, market drawdown, and related macro or stress states.

The paper follows the same empirical discipline used in the prior growth--defensive timing and cash-overlay allocation studies \cite{xiong2026growthdefensive,xiong2026cashoverlay}. Statistical screens are used to define the signal universe. Once a signal passes into the formal universe, it is not dynamically re-screened inside the rolling OOS loop. Walk-forward validation re-selects only the combination weights on already admitted signal groups. This separation is important. It prevents the strategy from changing its economic vocabulary during the OOS period while still allowing the relative importance of admitted signal families to adapt.

The final method is RGRR. Its implementation label in the archived CSV files is \code{filtered_ix3_top2}, but the paper uses RGRR as the method name because the core idea is economic rather than a code detail. RGRR uses one globally screened main effect, nine second-order interactions, and two third-order interactions selected by an additional incremental OOS filter. Earlier variants that used all statistically screened third-order interactions are retained as diagnostics, but they are not the final strategy. RGRR improves Sharpe versus QQQ in every tested interval, but it does not dominate QQQ on CAGR during strong QQQ-led markets. This is a meaningful but bounded result.

\begin{table}[H]
\centering
\caption{Structural Correspondence with the Cash-Overlay Study}
\label{tab:cash-overlay-correspondence}
\scriptsize
\begin{tabular}{p{4.5cm}p{4.7cm}p{4.8cm}}
\toprule
Cash-Overlay Study Block & Current Paper Counterpart & Reason for the Mapping\\
\midrule
Static risky sleeve and cash return & QQQ--DIA relative allocation problem & The asset-allocation object changes from risky sleeve versus cash to QQQ versus DIA.\\
Raw state variables & Relative, rate, volatility, credit, and drawdown states & Both papers begin from observable state variables before building scores.\\
Statistical screening and validation design & Global HAC screen plus rolling OOS validation & The same distinction is kept between statistical evidence and tradable OOS evidence.\\
Baseline annual diagnostics & Static QQQ/DIA baselines, factor attribution, annual matrix & The baseline difficulty is different: QQQ is a strong unconditional benchmark.\\
Policy grids and selection objectives & Fixed mapping plus lambda grid & The present allocation problem has no cash weight, but it still needs a score-to-weight grid and selection objective.\\
Slow-tail compensation filter & Lower-order relative relief filter & The lower-order layer captures persistent relative and macro-relief states.\\
V-shape crash-brake filter & Third-order conditional relief filter & The high-order layer is a conditional state filter, not a broad all-signal overlay.\\
Max-cash combination & Relief-Gated Relative Rotation & The final policy combines admitted signal groups into one deployable allocation rule.\\
Rolling and expanding start-date sensitivity & Rolling formal result plus expanding diagnostic & Rolling is the final validation; expanding is retained for format and robustness.\\
Appendix screening tables & Full interaction screen, IX3 incremental table, annual matrix & The appendix preserves the experiment audit trail.\\
\bottomrule
\end{tabular}
\end{table}

\section{Related Literature}

The study is related to three strands of empirical asset-allocation research. First, it follows the return-predictability literature in treating in-sample significance with caution. Goyal and Welch \cite{goyal2008comprehensive} and Campbell and Thompson \cite{campbell2008predicting} show that many predictive variables fail out of sample. This motivates the distinction between a statistical screen and a tradable claim. A variable can be statistically useful for organizing states without necessarily improving a portfolio after costs.

Second, the policy design is close to parametric portfolio rules. Brandt, Santa-Clara, and Valkanov \cite{brandt2009parametric} link portfolio weights directly to state variables. The present paper uses the same broad idea but restricts the action space to a two-ETF sleeve. The weight mapping is intentionally simple: an oriented score is standardized, transformed through a bounded nonlinear map, lagged, and smoothed.

Third, the validation problem is a data-snooping problem. Multiple signals, interactions, horizons, and mapping parameters create a search space. White's Reality Check \cite{white2000reality}, Hansen's SPA test \cite{hansen2005superior}, and the Deflated Sharpe Ratio of Bailey and Lopez de Prado \cite{bailey2014probability} are therefore relevant future confirmation tools. The current paper does not claim a final multiple-testing-adjusted anomaly. It reports a structured strategy experiment with explicit walk-forward tests, signal admission rules, and a narrower interpretation of the final result.

The factor-attribution table also uses the Fama--French and momentum tradition \cite{fama1993common,fama2015five,carhart1997persistence}. The relative return between QQQ and DIA has a strong growth-versus-value structure: QQQ has higher market beta and negative value and investment loadings, while DIA has more value, profitability, and investment exposure.

\section{QQQ--DIA Relative Allocation Problem}

Let $r^Q_t$ denote the daily return of QQQ and $r^D_t$ denote the daily return of DIA. The strategy holds a QQQ weight $w^Q_t \in [0,1]$ and a DIA weight $1-w^Q_t$. The net daily portfolio return is
\begin{equation}
r^P_t = w^Q_t r^Q_t + (1-w^Q_t)r^D_t - c \cdot TO_t,
\end{equation}
where $c=10$ basis points and turnover is measured as
\begin{equation}
TO_t = 2\left|w^Q_t-w^Q_{t-1}\right|.
\end{equation}
The factor of two reflects that shifting from one ETF to the other requires selling one sleeve and buying the other sleeve. All reported strategy returns are net of this cost convention. All reported Sharpe ratios use raw daily returns rather than excess returns over the risk-free rate.

The screening target is future QQQ--DIA relative return. For horizon $h$, the diagnostic target is the forward relative return between QQQ and DIA over $h$ trading days. The main horizons are 21, 63, and 126 days, with 63 and 126 days emphasized because the strategy is a tactical allocation rule rather than an intraday or very short-horizon rule.

\begin{table}[H]
\centering
\caption{Data Coverage and Portfolio Definitions}
\label{tab:data-coverage}
\scriptsize
\begin{tabular}{p{3.1cm}p{5.4cm}p{5.4cm}}
\toprule
Object & Definition & Sample or Role\\
\midrule
QQQ & Nasdaq-100 ETF, adjusted daily return from local Moomoo data. & Growth/technology-heavy equity sleeve. Return coverage: 2006-06-22--2026-07-02, 5018 daily returns.\\
DIA & Dow Jones Industrial Average ETF, adjusted daily return from local Moomoo data. & Large-cap industrial/value-tilted equity sleeve. Return coverage: 2006-06-22--2026-07-02, 5018 daily returns.\\
Strategy return & $w^Q_t r^Q_t+(1-w^Q_t)r^D_t$ minus turnover cost. & Tradable two-ETF relative allocation.\\
Baselines & 100\% QQQ, 100\% DIA, and 50/50 QQQ--DIA. & Benchmarks for return, Sharpe, drawdown, and final wealth.\\
Main OOS starts & 2008-07-25, 2018-06-28, 2020-01-02, 2022-01-03. & 2008 is a long-sample robustness start; 2018, 2020, and 2022 are the main screen windows for incremental third-order tests.\\
\bottomrule
\end{tabular}
\end{table}

\subsection{Raw State Variables}

The feature panel contains relative QQQ--DIA states and common macro or stress states. The relative states are designed to capture QQQ leadership, reversal pressure, and relative drawdown. The macro and stress states are used to identify when relative reversal is more likely to matter.

\begin{table}[H]
\centering
\caption{Raw State Variables Used in the RGRR Signal Layer}
\label{tab:raw-state-variables}
\scriptsize
\begin{tabular}{p{3.1cm}p{5.8cm}p{5.1cm}}
\toprule
State & Construction & Economic Role\\
\midrule
\code{rel_mom126} & Expanding-standardized 126-day QQQ--DIA relative momentum. & Captures persistent QQQ leadership or lagging.\\
\code{rel_reversal} & Expanding-standardized depth of the QQQ/DIA relative-ratio drawdown. & Captures reversal pressure after QQQ relative strength has weakened.\\
\code{rate_relief} & Negative expanding-standardized 21-day change in the 10-year Treasury yield. & Measures falling-rate relief.\\
\code{spy_drawdown} & Expanding-standardized depth of SPY drawdown. & Measures broad equity stress and rebound setup.\\
\code{high_vix} & Expanding-standardized VIX percentile. & Measures volatility stress.\\
\code{low_vix} & Negative of the high-VIX state. & Measures calm volatility conditions.\\
\code{vix_relief} & Negative expanding-standardized 21-day VIX change. & Measures easing volatility pressure.\\
\code{credit_relief} & Expanding-standardized 21-day HYG--SHY relative return. & Measures improving credit appetite.\\
\code{credit_stress} & Negative of the credit-relief state. & Measures credit risk-off pressure.\\
\bottomrule
\end{tabular}
\end{table}

All continuous state variables use expanding standardization with a 252-trading-day warmup. Interaction terms are formed from standardized states and are then expanding-standardized again. This keeps the score scale stable across main effects, second-order interactions, and third-order interactions.

\subsection{Univariate State Diagnostics}

Before forming interactions, the state variables were inspected through simple 63-day forward QQQ--DIA relative-return buckets. These tables are not used as the final admission rule because the final screen uses HAC regressions and walk-forward portfolio tests. They are included to show why the initial intuition required a conditional design rather than a one-variable rule.

\begin{table}[H]
\centering
\caption{Selected State Buckets and Forward 63-Day QQQ--DIA Relative Return}
\label{tab:state-quantile-diagnostics}
\scriptsize
\begin{tabular}{p{3.0cm}lrrp{5.4cm}}
\toprule
State & Bucket & Mean & Median & Interpretation\\
\midrule
\code{rel_trailing_126d_z} & Q1 & \pct{2.87} & \pct{4.30} & Weak prior QQQ relative momentum is followed by stronger future QQQ relative return.\\
\code{rel_trailing_126d_z} & Q3 & \pct{1.74} & \pct{1.66} & Middle bucket has moderate future relative return.\\
\code{rel_trailing_126d_z} & Q5 & \pct{0.52} & \pct{0.96} & Strong prior QQQ leadership does not by itself predict further strong QQQ relative return.\\
\code{rel_drawdown_depth} & Q1 & \pct{0.58} & \pct{0.73} & Shallow relative drawdown contains little reversal pressure.\\
\code{rel_drawdown_depth} & Q3 & \pct{0.00} & \pct{0.92} & Middle bucket is weak and noisy.\\
\code{rel_drawdown_depth} & Q5 & \pct{3.83} & \pct{4.56} & Deep QQQ relative drawdown has the clearest positive forward relative-return bucket.\\
\code{rate_relief} & Q1 & \pct{1.21} & \pct{1.72} & Low rate relief is not enough for a strong QQQ relative state.\\
\code{rate_relief} & Q3 & \pct{1.76} & \pct{1.75} & Middle relief is modestly positive.\\
\code{rate_relief} & Q5 & \pct{2.05} & \pct{2.21} & Strong rate relief has the best univariate forward bucket.\\
\code{spy_drawdown_depth} & Q1 & \pct{1.96} & \pct{1.71} & Low market drawdown is not the main catch-up state.\\
\code{spy_drawdown_depth} & Q3 & \pct{0.94} & \pct{1.04} & Middle drawdown is weaker.\\
\code{spy_drawdown_depth} & Q5 & \pct{3.00} & \pct{3.81} & Deep broad-market drawdown raises the value of conditional relief terms.\\
\bottomrule
\end{tabular}
\end{table}

The bucket evidence is consistent with the final RGRR design. Prior QQQ leadership alone is not enough; the highest \code{rel_trailing_126d_z} bucket has the lowest mean forward relative return among the displayed momentum buckets. By contrast, deep relative drawdown and strong rate relief have clearer positive buckets. This explains why the final method keeps relative reversal and rate relief in interactions rather than building a simple rule that fades QQQ whenever it has led DIA.

\section{Statistical Screening and Validation Design}

The screen begins with horizon-specific regressions of future QQQ--DIA relative return on each candidate signal:
\begin{equation}
R^{Q-D}_{t,t+h}=\alpha_h+\beta_h x_t+\epsilon_{t,h},\qquad h\in\{21,63,126\}.
\end{equation}
Because the future-return labels overlap, the reported $t$-statistics use Newey--West/HAC standard errors. For each candidate, the best horizon is retained if the absolute HAC $t$-statistic is at least 2.0. Highly correlated candidates are de-duplicated using an absolute correlation threshold of 0.95.

The sign of the coefficient determines the orientation. If the coefficient is positive, the candidate enters the score with orientation $+1$; if the coefficient is negative, it enters with orientation $-1$. Thus a positive oriented score means a state that historically pointed toward stronger future QQQ--DIA relative return. The orientation is fixed after the global screen.

The formal validation layer is walk-forward OOS. The rolling window uses 756 training trading days and 63-day test blocks. For each block, the strategy re-selects only the lambdas on the already admitted signal groups. It does not re-screen the signal universe. The selection objective uses training performance and includes a turnover penalty above 300\% annualized turnover. The mapping parameters are fixed globally as strategy attributes rather than re-selected in each rolling block.

\section{Baseline Annual Diagnostics}

The static baseline already favors QQQ over the full 2008--2026 sample. From 2008-07-25 to 2026-07-02, 100\% QQQ has the highest CAGR and Sharpe among the static QQQ/DIA mixes. This makes the strategy test difficult: a dynamic rule must either beat QQQ on risk-adjusted return or justify lower CAGR through materially better drawdown control.

\begin{table}[H]
\centering
\caption{Static Baselines, 2008-07-25--2026-07-02}
\label{tab:static-baselines}
\scriptsize
\begin{tabular}{lrrrr}
\toprule
Portfolio & CAGR & Sharpe & Max DD & Final Wealth\\
\midrule
100\% DIA & \pct{11.65} & 0.66 & \pct{-44.54} & 7.13\\
50/50 QQQ--DIA & \pct{14.62} & 0.78 & \pct{-45.53} & 11.38\\
100\% QQQ & \pct{17.17} & 0.82 & \pct{-47.83} & 16.86\\
\bottomrule
\end{tabular}
\end{table}

Annual diagnostics show why the RGRR claim must be framed as a relative-allocation and risk-adjusted-return claim rather than as an unconditional QQQ replacement. In strong QQQ years such as 2019, 2020, and 2023, a dynamic rule that sometimes rotates toward DIA will usually trail 100\% QQQ on calendar-year return. In drawdown-heavy or style-shifting years, especially 2022, the same ability to reduce QQQ exposure becomes valuable.

\begin{table}[H]
\centering
\caption{Annual Return / Maximum Drawdown on the 2008-Start Rolling Path}
\label{tab:annual-baseline}
\begingroup
\tiny
\setlength{\tabcolsep}{2pt}
\resizebox{\textwidth}{!}{%
\begin{tabular}{lrrrrr}
\toprule
Year & RGRR & 100\% QQQ & 100\% DIA & 50/50 QQQ--DIA & Avg. RGRR QQQ\\
\midrule
2016 & \pct{13.72} / \pct{-11.06} & \pct{7.22} / \pct{-12.12} & \pct{16.39} / \pct{-8.47} & \pct{11.82} / \pct{-10.16} & \pct{53.30}\\
2017 & \pct{29.03} / \pct{-2.57} & \pct{32.94} / \pct{-4.91} & \pct{28.23} / \pct{-3.21} & \pct{30.68} / \pct{-2.11} & \pct{41.59}\\
2018 & \pct{-4.56} / \pct{-19.43} & \pct{-0.13} / \pct{-22.89} & \pct{-3.76} / \pct{-18.15} & \pct{-1.81} / \pct{-20.41} & \pct{29.92}\\
2019 & \pct{31.41} / \pct{-9.00} & \pct{39.16} / \pct{-11.03} & \pct{25.04} / \pct{-6.70} & \pct{32.02} / \pct{-8.59} & \pct{55.48}\\
2020 & \pct{28.99} / \pct{-29.25} & \pct{48.42} / \pct{-28.65} & \pct{9.54} / \pct{-36.71} & \pct{28.04} / \pct{-32.28} & \pct{45.04}\\
2021 & \pct{24.31} / \pct{-5.16} & \pct{27.42} / \pct{-10.86} & \pct{20.83} / \pct{-6.37} & \pct{24.42} / \pct{-5.64} & \pct{38.15}\\
2022 & \pct{-15.23} / \pct{-23.61} & \pct{-32.68} / \pct{-34.83} & \pct{-7.04} / \pct{-20.76} & \pct{-20.60} / \pct{-26.80} & \pct{30.71}\\
2023 & \pct{37.61} / \pct{-9.20} & \pct{55.40} / \pct{-10.78} & \pct{16.16} / \pct{-8.56} & \pct{34.64} / \pct{-9.10} & \pct{63.19}\\
2024 & \pct{21.40} / \pct{-9.98} & \pct{25.75} / \pct{-13.56} & \pct{14.82} / \pct{-6.05} & \pct{20.46} / \pct{-8.93} & \pct{61.75}\\
2025 & \pct{18.06} / \pct{-20.23} & \pct{20.96} / \pct{-22.77} & \pct{14.83} / \pct{-15.83} & \pct{18.10} / \pct{-19.15} & \pct{60.27}\\
2026 & \pct{29.00} / \pct{-9.66} & \pct{35.52} / \pct{-11.72} & \pct{22.67} / \pct{-9.76} & \pct{29.39} / \pct{-9.66} & \pct{54.46}\\
\bottomrule
\end{tabular}
}
\endgroup
\end{table}

\begin{figure}[H]
\centering
\includegraphics[width=0.98\textwidth]{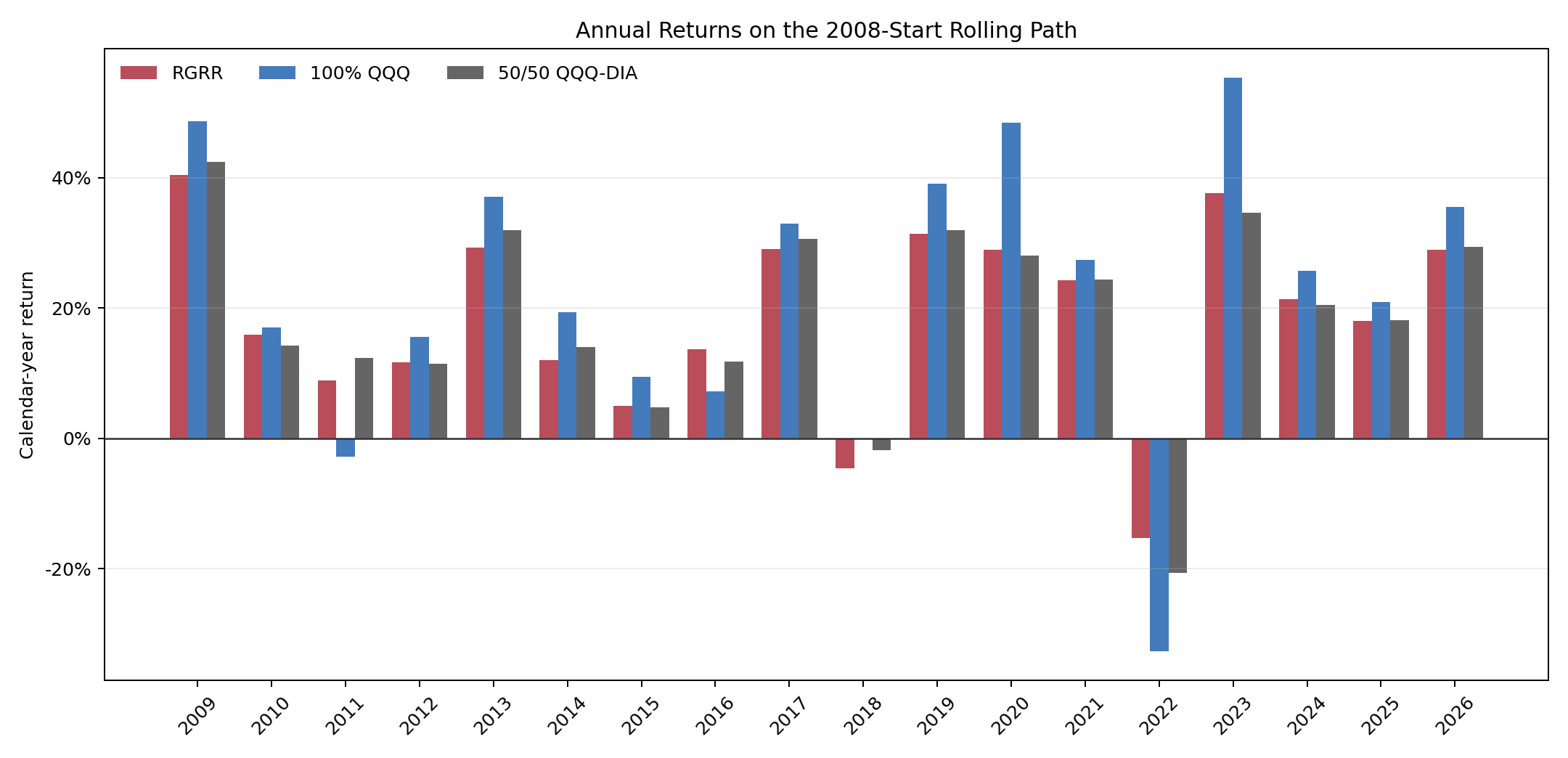}
\caption{Annual Returns on the 2008-Start Rolling Path}
\label{fig:yearly-returns}
\end{figure}

The annual pattern is consistent with the final interpretation. RGRR's largest calendar-year advantage versus QQQ occurs in 2022, when it loses -15.23\% versus -32.68\% for QQQ. It also reduces drawdown in several non-crisis years, but it does not fully participate in the strongest QQQ rebounds. This is why the formal claim later focuses on Sharpe improvement and 50/50 benchmark improvement, not on unconditional dominance over QQQ.

\subsection{Factor Attribution}

The factor attribution confirms that the relative asset pair is economically meaningful rather than arbitrary. QQQ has a higher market beta and negative value/investment loadings, while DIA has lower market beta and positive value, profitability, and investment loadings. The relative QQQ--DIA return has a positive annualized intercept in the six-factor regression, though its Newey--West $t$-statistic is below the conventional 2.0 threshold.

\begin{table}[H]
\centering
\caption{Fama--French Five-Factor plus Momentum Attribution}
\label{tab:ff-attribution}
\tiny
\setlength{\tabcolsep}{3pt}
\begin{tabular}{lrrrrrrrrr}
\toprule
Portfolio & Alpha Ann. & Alpha $t$ & MKT & SMB & HML & RMW & CMA & MOM & Adj. $R^2$\\
\midrule
QQQ excess return & \pct{3.57} & 2.72 & 1.08 & -0.06 & -0.31 & 0.01 & -0.22 & 0.04 & 0.92\\
DIA excess return & \pct{-0.62} & -0.41 & 0.93 & -0.08 & 0.09 & 0.22 & 0.17 & -0.00 & 0.86\\
QQQ--DIA relative return & \pct{4.19} & 1.92 & 0.15 & 0.02 & -0.40 & -0.21 & -0.39 & 0.04 & 0.38\\
\bottomrule
\end{tabular}
\end{table}

\section{Policy Grids and Selection Objectives}

The final signal score is a weighted sum of admitted signal groups:
\begin{equation}
S_t = \lambda_m S^{main}_t+\lambda_2 S^{ix2}_t+\lambda_3 S^{ix3}_t.
\end{equation}
The lambdas are selected from
\begin{equation}
\lambda_m,\lambda_2,\lambda_3 \in \{0.25,0.50,0.75,1.00\}.
\end{equation}
For the lower-order base policy, $\lambda_3=0$. For the final policy, $\lambda_3$ applies only to the two third-order terms that pass the incremental OOS filter.

The policy grid is intentionally small. The final method does not estimate unconstrained coefficients for every signal. Instead, it groups admitted variables into economically interpretable blocks and searches over coarse positive lambdas. This prevents the OOS selector from turning into a high-dimensional regression engine. The grid has two layers:
\begin{enumerate}
\item A \emph{fixed mapping layer}, selected as a strategy attribute rather than re-estimated inside each rolling block.
\item A \emph{signal-combination layer}, where the rolling selector chooses lambdas on the admitted main, second-order, and filtered third-order signal groups.
\end{enumerate}

The position mapping is
\begin{equation}
\tilde{w}^Q_t=0.5+M\tanh\left(Z(S_t)/\tau\right),
\end{equation}
where $Z(S_t)$ is an expanding standardized score. The final fixed mapping is
\begin{equation}
M=0.50,\qquad \tau=0.75,\qquad \eta=0.05.
\end{equation}
The target weight is lagged by one trading day and smoothed:
\begin{equation}
w^Q_t=(1-\eta)w^Q_{t-1}+\eta\tilde{w}^Q_{t-1}.
\end{equation}
This lag avoids same-day lookahead, and the smoothing controls turnover. The fixed mapping was selected as a strategy attribute through the global sensitivity experiment. It is not re-selected inside the rolling loop.

\begin{table}[H]
\centering
\caption{RGRR Policy Grid Design}
\label{tab:grid-design}
\scriptsize
\begin{tabular}{p{3.2cm}p{5.2cm}p{5.6cm}}
\toprule
Layer & Grid or Fixed Values & Role\\
\midrule
Signal lambdas & $\lambda_m,\lambda_2,\lambda_3 \in \{0.25,0.50,0.75,1.00\}$ & Re-selected inside each walk-forward block using only the training window.\\
Lower-order base & $\lambda_3=0$ & Diagnostic base policy for testing whether third-order interactions add incremental OOS value.\\
Final RGRR stack & $\lambda_3>0$ only after the third-order incremental screen & Formal final strategy after the filtered third-order set is admitted.\\
Position range & $M=0.50$ & Allows QQQ weight to range from 0\% to 100\% before smoothing.\\
Score scale & $\tau=0.75$ & Controls sensitivity of the tanh weight map to the standardized score.\\
Smoothing speed & $\eta=0.05$ & Makes the mapping responsive but prevents one-day jumps to the raw target.\\
Selection objective & Training selection score minus penalty for annual turnover above 300\% & Keeps the OOS selector from choosing high-turnover variants simply because their in-sample score is marginally better.\\
\bottomrule
\end{tabular}
\end{table}

This design responds to a key methodological point from the experiment log. Sensitivity parameters such as $\tau$ and $\eta$ should not be dynamically re-selected every rolling window if they are part of the strategy's behavioral identity. They determine how aggressively the method turns a state estimate into a tradable position. The final paper therefore treats $M$, $\tau$, and $\eta$ as fixed attributes of RGRR, while allowing the relative weight on already admitted signal families to adapt by rolling training evidence.

\begin{table}[H]
\centering
\caption{Top Fixed-Mapping Sensitivity Rows by Rolling Sharpe}
\label{tab:mapping-sensitivity}
\tiny
\setlength{\tabcolsep}{3pt}
\begin{tabular}{lrrrrrr}
\toprule
Mapping & Max Tilt & $\tau$ & $\eta$ & Rolling CAGR & Rolling Sharpe & Rolling Max DD\\
\midrule
\code{tilt0p50_tau0p75_eta0p05} & 0.50 & 0.75 & 0.05 & \pct{17.32} & 0.91 & \pct{-29.52}\\
\code{tilt0p50_tau0p50_eta0p05} & 0.50 & 0.50 & 0.05 & \pct{17.19} & 0.91 & \pct{-29.40}\\
\code{tilt0p50_tau1p00_eta0p05} & 0.50 & 1.00 & 0.05 & \pct{17.34} & 0.91 & \pct{-29.54}\\
\code{tilt0p50_tau1p00_eta0p03} & 0.50 & 1.00 & 0.03 & \pct{17.36} & 0.91 & \pct{-30.51}\\
\code{tilt0p50_tau0p75_eta0p10} & 0.50 & 0.75 & 0.10 & \pct{17.12} & 0.90 & \pct{-28.75}\\
\bottomrule
\end{tabular}
\end{table}

The selected mapping is not the most conservative mapping. It is a relatively responsive mapping with full possible tilt between 0\% and 100\% QQQ, a moderate score scale, and slow daily adjustment. This fits the empirical interpretation: the signal needs enough sensitivity to react to relative-regime changes, while the daily update must still avoid extreme one-day turnover.

\section{Lower-Order Relative Relief Filter}

\subsection{Policy Construction}

The global screen selects only one main effect: \code{rate_relief}. This is a useful result because it means the formal main-effect layer is not a broad collection of loosely significant variables. It says that falling-rate relief, by itself, has the clearest global association with future QQQ--DIA relative return among the tested main effects.

The second-order layer contains nine interactions. These interactions combine relative reversal or momentum with rate, volatility, credit, and market-drawdown states. They are not dynamically chosen during OOS. They are admitted once by the global screen and then held fixed.

Let $\mathcal{M}$ be the admitted main-effect set and $\mathcal{I}_2$ be the admitted second-order set. Each signal has a fixed orientation $o_j \in \{-1,+1\}$ determined by the sign of its global HAC regression coefficient. The lower-order score is
\begin{align}
S^{main}_t &= Z\left(\frac{1}{|\mathcal{M}|}\sum_{j\in \mathcal{M}} o_j x_{j,t}\right),\\
S^{ix2}_t &= Z\left(\frac{1}{|\mathcal{I}_2|}\sum_{j\in \mathcal{I}_2} o_j x_{j,t}\right),
\end{align}
where $Z(\cdot)$ denotes expanding standardization. The lower-order base policy uses
\begin{equation}
S^{base}_t=\lambda_m S^{main}_t+\lambda_2 S^{ix2}_t.
\end{equation}
This base policy is important because the third-order experiment is judged relative to it. A third-order term is not admitted merely because it is statistically significant; it must add OOS value after the lower-order relief filter already exists.

\begin{table}[H]
\centering
\caption{Final Globally Screened Lower-Order Signal Set}
\label{tab:lower-order-signals}
\scriptsize
\begin{tabular}{p{5.6cm}ccr}
\toprule
Signal & Order & Horizon & HAC $t$\\
\midrule
\code{rate_relief} & main & 126 & 2.36\\
\code{vix_relief * credit_stress} & 2 & 126 & -3.62\\
\code{rel_mom126 * credit_relief} & 2 & 126 & -2.79\\
\code{rate_relief * low_vix} & 2 & 21 & -2.75\\
\code{rate_relief * credit_relief} & 2 & 126 & -2.69\\
\code{rel_reversal * high_vix} & 2 & 126 & -2.50\\
\code{rate_relief * spy_drawdown} & 2 & 126 & 2.48\\
\code{rel_reversal * rate_relief} & 2 & 63 & 2.47\\
\code{credit_relief * credit_stress} & 2 & 21 & -2.31\\
\code{rel_reversal * credit_relief} & 2 & 126 & 2.19\\
\bottomrule
\end{tabular}
\end{table}

The lower-order layer already contains the key economic vocabulary of the final strategy. DIA catch-up is not supported as a simple rule based only on QQQ having led DIA. Instead, relative reversal enters together with rate relief, high VIX, or credit relief. This is why the final interpretation is conditional: relative reversal matters, but mostly inside broader macro and stress states.

\begin{table}[H]
\centering
\caption{Rolling Main+IX2 Base Policy before Third-Order Admission}
\label{tab:base-main-ix2}
\scriptsize
\begin{tabular}{lrrrrrr}
\toprule
Period & CAGR & Sharpe & Max DD & Ann. Turnover & Avg. QQQ & Selected Configs\\
\midrule
2008 start & \pct{16.59} & 0.97 & \pct{-29.24} & \pct{500.64} & \pct{46.46} & 10\\
2018 OOS & \pct{17.65} & 0.93 & \pct{-29.61} & \pct{449.11} & \pct{44.11} & 5\\
2020 OOS & \pct{17.89} & 0.91 & \pct{-29.61} & \pct{425.30} & \pct{44.56} & 5\\
2022 OOS & \pct{14.19} & 0.84 & \pct{-23.78} & \pct{388.50} & \pct{45.48} & 5\\
\bottomrule
\end{tabular}
\end{table}

Table~\ref{tab:base-main-ix2} shows that the lower-order base already improves risk-adjusted behavior relative to static 50/50, but it leaves room for incremental improvement. The third-order filter is therefore not used to discover the whole strategy from scratch. It is used only to decide whether a small number of conditional relief terms can improve the already-screened lower-order base.

\section{Third-Order Conditional Relief Filter}

\subsection{Policy Construction}

The first all-signal version allowed every statistically screened third-order interaction into the policy. That was too permissive. The final paper therefore adds a second gate for third-order terms. A third-order interaction must satisfy all of the following:
\begin{enumerate}
\item pass the global HAC screen and correlation de-duplication;
\item improve rolling OOS Sharpe versus the \code{main+ix2} base policy on average across the 2018, 2020, and 2022 OOS screen periods;
\item have positive Sharpe delta in at least two of the three screen periods;
\item survive economic-family de-duplication, with a maximum of five third-order terms allowed.
\end{enumerate}

This rule is stricter than the lower-order rule because third-order terms have higher overfitting risk. It also follows the principle used in the prior growth--defensive timing and cash-overlay studies that complex interactions should be kept only when they add portfolio evidence beyond simpler components \cite{xiong2026growthdefensive,xiong2026cashoverlay}.

Let $\mathcal{I}_3^\star$ be the retained third-order set. The conditional relief score is
\begin{equation}
S^{ix3}_t=Z\left(\frac{1}{|\mathcal{I}_3^\star|}\sum_{j\in \mathcal{I}_3^\star} o_j x_{j,t}\right).
\end{equation}
The final RGRR score is then
\begin{equation}
S^{RGRR}_t=\lambda_m S^{main}_t+\lambda_2 S^{ix2}_t+\lambda_3 S^{ix3}_t.
\end{equation}
If no third-order term passed the incremental screen, the correct formal method would have been the lower-order base. The cap of five terms is therefore only a maximum complexity budget; it is not a target to fill.

\begin{table}[H]
\centering
\caption{Retained Third-Order Interactions after Incremental OOS Filter}
\label{tab:ix3-retained}
\scriptsize
\resizebox{\textwidth}{!}{%
\begin{tabular}{p{5.7cm}p{3.2cm}rrrr}
\toprule
Interaction & Economic Family & Mean $\Delta$ Sharpe & Positive Periods & Mean $\Delta$ CAGR & HAC $|t|$\\
\midrule
\code{rel_mom126 * rel_reversal * rate_relief} & rate + relative & 0.019 & 3 & \pct{0.50} & 2.78\\
\code{rel_reversal * rate_relief * spy_drawdown} & market drawdown + rate + relative & 0.000 & 2 & \pct{0.14} & 4.09\\
\bottomrule
\end{tabular}
}
\end{table}

Only two of the 26 statistically screened third-order interactions survive this additional test. This is the main methodological change from the earlier full-IX3 experiment. The retained terms both include \code{rel_reversal} and \code{rate_relief}; one also includes prior relative momentum, and the other includes broad market drawdown. This supports a more nuanced version of the DIA catch-up idea: reversal pressure becomes tradable only when rate relief and market context make the relative move economically coherent.

\begin{table}[H]
\centering
\caption{Filtered Third-Order Set Test versus Lower-Order Base}
\label{tab:ix3-set-test-main}
\scriptsize
\begin{tabular}{lrrrrr}
\toprule
Policy & IX3 Terms & Mean $\Delta$ Sharpe & Positive Periods & Mean $\Delta$ CAGR & Mean Turnover\\
\midrule
Top 2 retained set & 2 & 0.02 & 3 & \pct{0.83} & \pct{402.36}\\
Top 1 retained set & 1 & 0.02 & 3 & \pct{0.50} & \pct{364.21}\\
\bottomrule
\end{tabular}
\end{table}

The one-term and two-term filtered sets both improve mean OOS Sharpe versus the base. The two-term set is selected because it has the higher mean incremental CAGR and remains positive across the three screen windows. This is a portfolio-level selection, not a claim that the second term is strong on its own in every window; its individual mean Sharpe contribution is close to zero but positive under the stated screen and complementary to the first term.

\begin{figure}[H]
\centering
\includegraphics[width=0.98\textwidth]{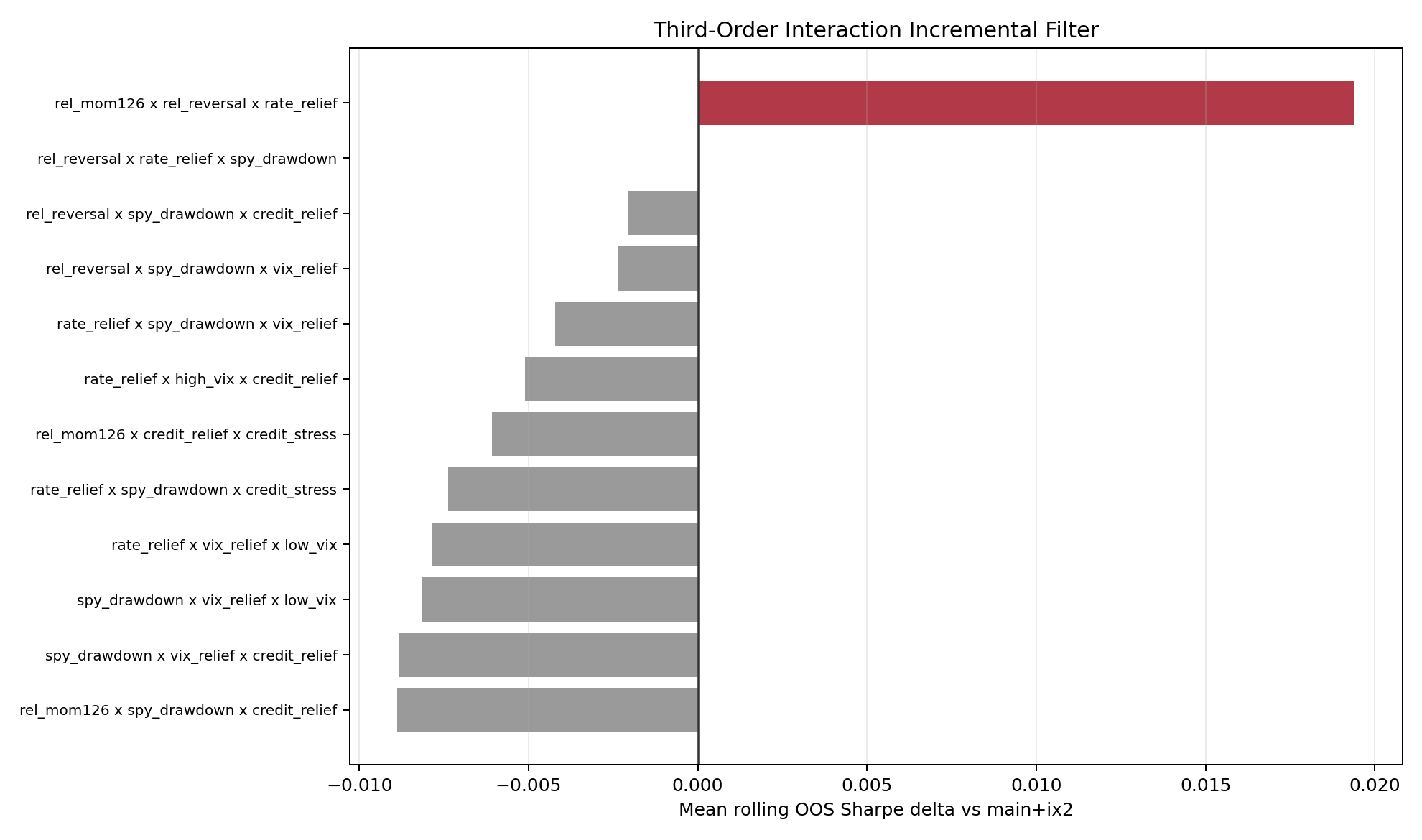}
\caption{Third-Order Interaction Incremental OOS Sharpe Filter}
\label{fig:ix3-incremental}
\end{figure}

Figure~\ref{fig:ix3-incremental} shows why the final strategy keeps only two third-order interactions. Many terms have strong global HAC statistics, but their incremental rolling OOS Sharpe contribution is negative once the main+ix2 policy is already present. These terms remain useful diagnostics of state structure, but they are not formal trading signals.

\section{Experiment Lineage and Rejected Alternatives}

The final RGRR design is the result of several rejected alternatives. This section is included because the experimental path matters for interpretation. A strategy can look attractive if the final table is shown without the earlier versions that failed a stricter rule. The research log therefore records the main design revisions:
\begin{enumerate}
\item \emph{Formal all-signals}: all statistically screened main, second-order, and third-order interactions were allowed into the score. This was too permissive because many third-order terms were statistically strong but portfolio-redundant.
\item \emph{Main+IX2 with fixed mapping}: the strategy removed the full third-order layer and fixed the mapping parameters as strategy attributes. This improved interpretability and reduced the risk of high-order overfitting.
\item \emph{Turnover-penalized main+IX2}: the rolling selector added an explicit penalty when training turnover exceeded 300\% annualized. This became the lower-order base for the final incremental IX3 test.
\item \emph{RGRR}: only third-order terms that improved rolling OOS Sharpe versus the lower-order base were admitted back into the final signal stack.
\end{enumerate}

\begin{table}[H]
\centering
\caption{Experiment Lineage: Rolling Results by Design Stage}
\label{tab:experiment-lineage}
\begingroup
\tiny
\setlength{\tabcolsep}{2pt}
\resizebox{\textwidth}{!}{%
\begin{tabular}{llrrrrr}
\toprule
Period & Design Stage & CAGR & Sharpe & Max DD & Ann. Turnover & Avg. QQQ\\
\midrule
2008 start & All statistically screened signals & \pct{16.21} & 0.95 & \pct{-29.28} & \pct{441.62} & \pct{46.97}\\
2008 start & Main+IX2, standard selector & \pct{16.55} & 0.96 & \pct{-29.24} & \pct{510.57} & \pct{46.53}\\
2008 start & Main+IX2, turnover-penalized selector & \pct{16.59} & 0.97 & \pct{-29.24} & \pct{500.64} & \pct{46.46}\\
2008 start & RGRR & \pct{16.85} & 0.97 & \pct{-29.25} & \pct{505.89} & \pct{48.83}\\
2018 OOS & All statistically screened signals & \pct{17.32} & 0.91 & \pct{-29.52} & \pct{321.00} & \pct{43.60}\\
2018 OOS & Main+IX2, standard selector & \pct{17.44} & 0.92 & \pct{-29.61} & \pct{476.08} & \pct{44.15}\\
2018 OOS & Main+IX2, turnover-penalized selector & \pct{17.65} & 0.93 & \pct{-29.61} & \pct{449.11} & \pct{44.11}\\
2018 OOS & RGRR & \pct{18.33} & 0.94 & \pct{-29.41} & \pct{444.60} & \pct{49.38}\\
2020 OOS & All statistically screened signals & \pct{17.42} & 0.89 & \pct{-29.43} & \pct{267.18} & \pct{43.35}\\
2020 OOS & Main+IX2, standard selector & \pct{17.70} & 0.90 & \pct{-29.61} & \pct{452.97} & \pct{44.59}\\
2020 OOS & Main+IX2, turnover-penalized selector & \pct{17.89} & 0.91 & \pct{-29.61} & \pct{425.30} & \pct{44.56}\\
2020 OOS & RGRR & \pct{18.69} & 0.93 & \pct{-29.41} & \pct{408.16} & \pct{50.37}\\
2022 OOS & All statistically screened signals & \pct{13.59} & 0.82 & \pct{-23.99} & \pct{257.71} & \pct{43.34}\\
2022 OOS & Main+IX2, standard selector & \pct{13.99} & 0.83 & \pct{-23.78} & \pct{424.68} & \pct{45.59}\\
2022 OOS & Main+IX2, turnover-penalized selector & \pct{14.19} & 0.84 & \pct{-23.78} & \pct{388.50} & \pct{45.48}\\
2022 OOS & RGRR & \pct{15.19} & 0.87 & \pct{-23.62} & \pct{354.33} & \pct{54.03}\\
\bottomrule
\end{tabular}
}
\endgroup
\end{table}

Table~\ref{tab:experiment-lineage} clarifies why the final method is not the earlier all-signal policy. The all-signal version often had lower turnover because it selected different lambdas and weight paths, but it did not produce the best risk-adjusted result. The main+IX2 base improved interpretability, and the filtered IX3 layer added back only the small set of high-order conditions that improved OOS Sharpe. This experimental lineage is a key part of the final methodology.

\section{Relief-Gated Relative Rotation}

RGRR is the final combined method. Its archived implementation label is \code{filtered_ix3_top2}, but the method name used in the paper is Relief-Gated Relative Rotation. Its formal signal universe is:
\begin{equation}
\text{1 main effect} + \text{9 second-order interactions} + \text{2 third-order interactions}.
\end{equation}
The rolling validation loop re-selects lambdas on these three groups. It does not re-screen the signal list and does not change the fixed mapping parameters.

\begin{table}[H]
\centering
\caption{Final Rolling OOS Results for RGRR}
\label{tab:final-oos}
\scriptsize
\begin{tabular}{lrrrrrr}
\toprule
Period & CAGR & Sharpe & Max DD & Ann. Turnover & Avg. QQQ & Final Wealth\\
\midrule
2008 start, actual 2009-07-27 & \pct{16.85} & 0.97 & \pct{-29.25} & \pct{505.89} & \pct{48.83} & 13.73\\
2018 OOS & \pct{18.33} & 0.94 & \pct{-29.41} & \pct{444.60} & \pct{49.38} & 3.84\\
2020 OOS & \pct{18.69} & 0.93 & \pct{-29.41} & \pct{408.16} & \pct{50.37} & 3.03\\
2022 OOS & \pct{15.19} & 0.87 & \pct{-23.62} & \pct{354.33} & \pct{54.03} & 1.88\\
\bottomrule
\end{tabular}
\end{table}

The final OOS results have a stable shape. Sharpe is above the QQQ baseline in every aligned period, but CAGR is not. This is economically reasonable because QQQ has a very high unconditional return in the sample. RGRR improves the balance between QQQ and DIA rather than replacing QQQ as a high-growth benchmark.

\begin{figure}[H]
\centering
\includegraphics[width=0.98\textwidth]{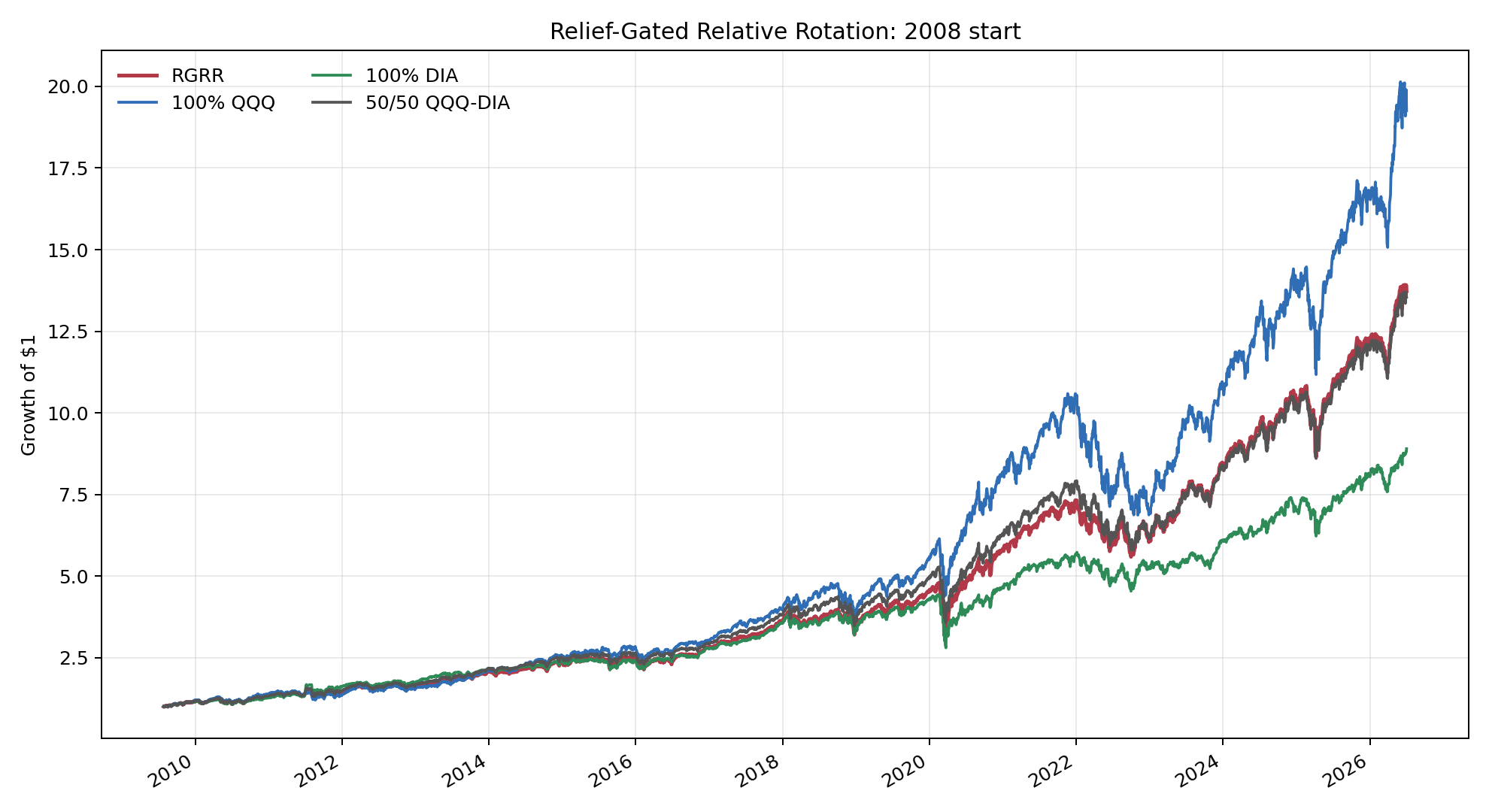}
\caption{RGRR and Baselines, Long 2008-Start Robustness Window}
\label{fig:final-2008-equity}
\end{figure}

The long-sample robustness window is the hardest comparison against QQQ because it includes an extended technology-led bull market. In Figure~\ref{fig:final-2008-equity}, RGRR's final wealth is close to the 50/50 benchmark and below QQQ, while its Sharpe is slightly above both. This supports the paper's narrow claim: RGRR is a risk-adjusted relative rotation rule, not a mechanism that should be expected to dominate QQQ terminal wealth in every growth-led sample.

\begin{figure}[H]
\centering
\includegraphics[width=0.98\textwidth]{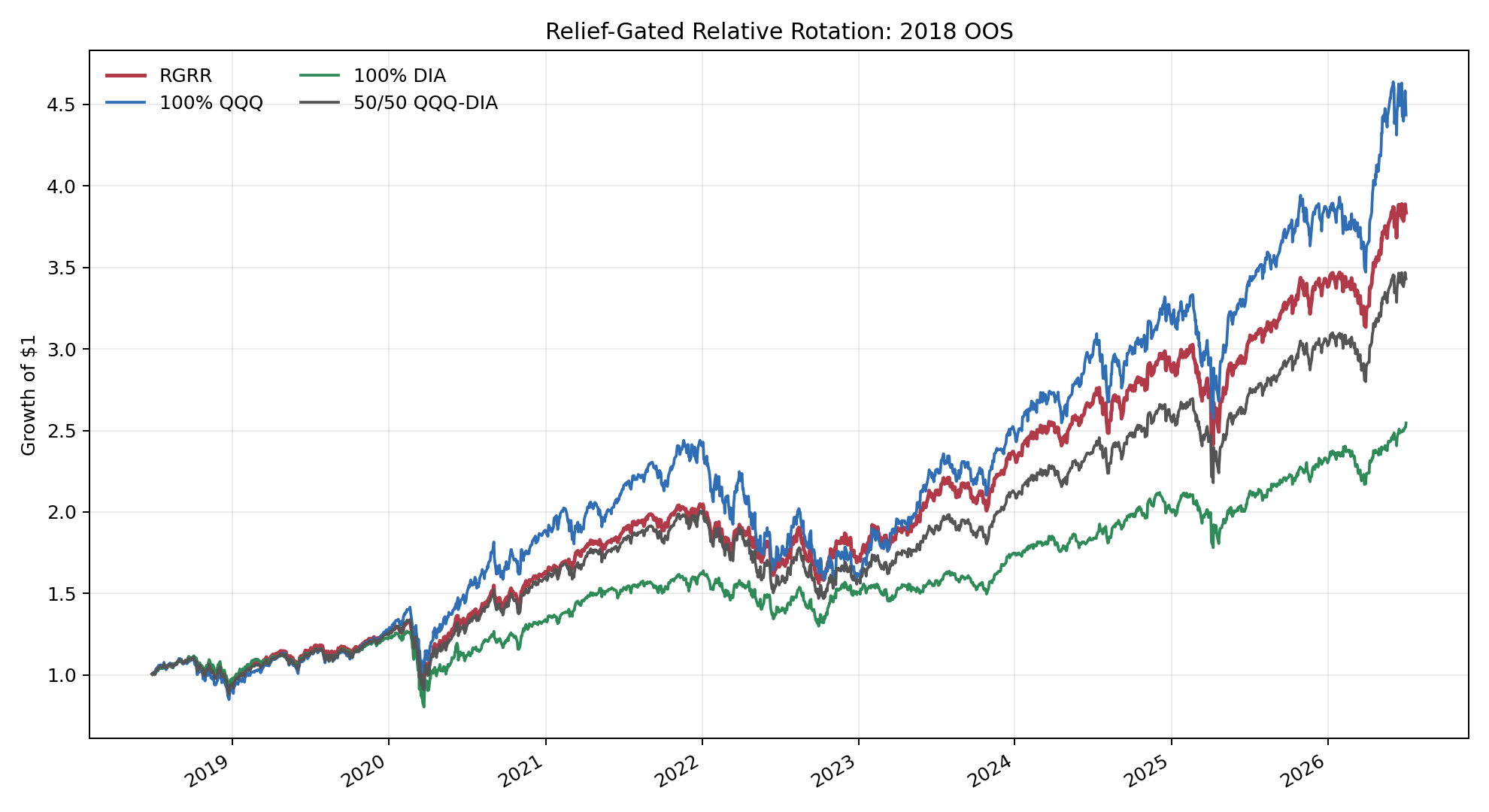}
\caption{RGRR and Baselines, 2018 OOS}
\label{fig:final-2018-equity}
\end{figure}

Figure~\ref{fig:final-2018-equity} shows the main trade-off in the longest formal OOS window. RGRR trails 100\% QQQ in terminal wealth but does so with a higher Sharpe and lower maximum drawdown. It also beats 50/50 QQQ--DIA on CAGR, Sharpe, drawdown, and final wealth.

\begin{figure}[H]
\centering
\includegraphics[width=0.98\textwidth]{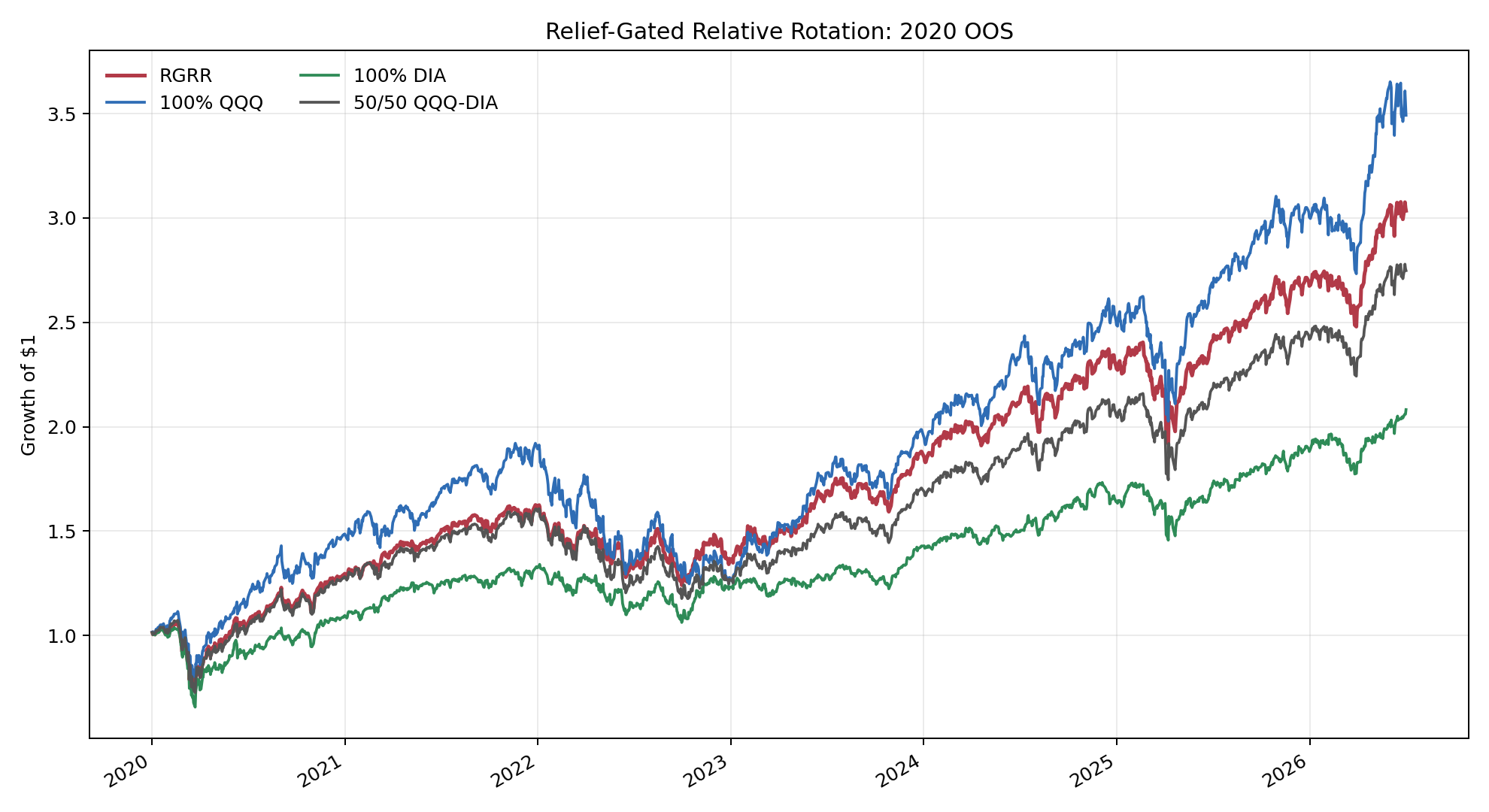}
\caption{RGRR and Baselines, 2020 OOS}
\label{fig:final-2020-equity}
\end{figure}

The 2020 OOS window illustrates the cost of rotating away from QQQ during an exceptional technology-led recovery. RGRR still improves Sharpe and drawdown versus QQQ, but it gives up CAGR. This is an important negative result because it prevents the method from being overstated as a universal QQQ outperformance rule.

\begin{figure}[H]
\centering
\includegraphics[width=0.98\textwidth]{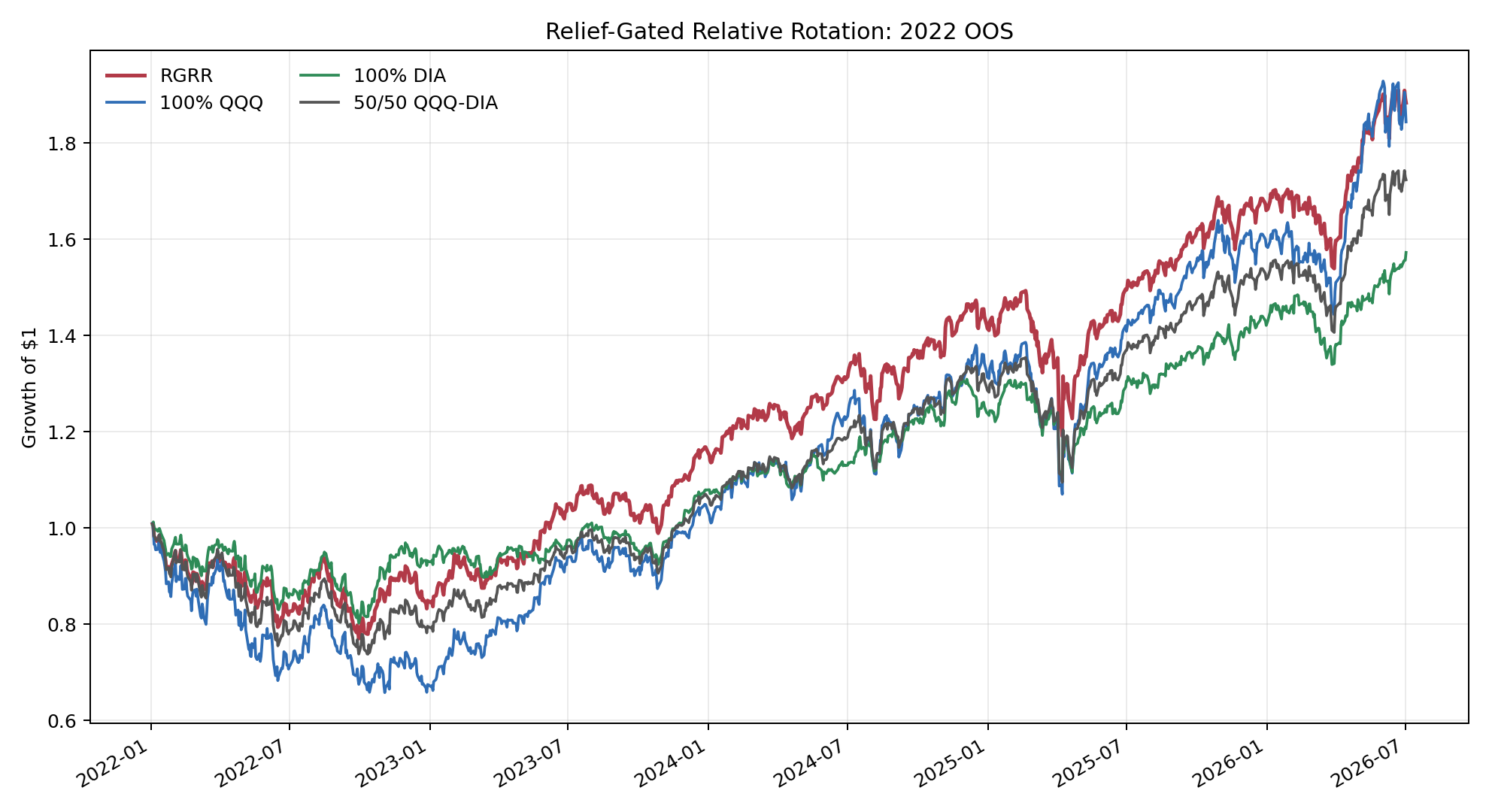}
\caption{RGRR and Baselines, 2022 OOS}
\label{fig:final-2022-equity}
\end{figure}

The 2022 OOS window is the strongest result relative to QQQ. In Figure~\ref{fig:final-2022-equity}, RGRR improves both CAGR and Sharpe versus QQQ and materially reduces the drawdown. It does not beat 100\% DIA on maximum drawdown, because DIA is the more defensive ETF in this window; however, it earns a higher CAGR and Sharpe than DIA.

\begin{figure}[H]
\centering
\begin{subfigure}{0.98\textwidth}
\includegraphics[width=\textwidth]{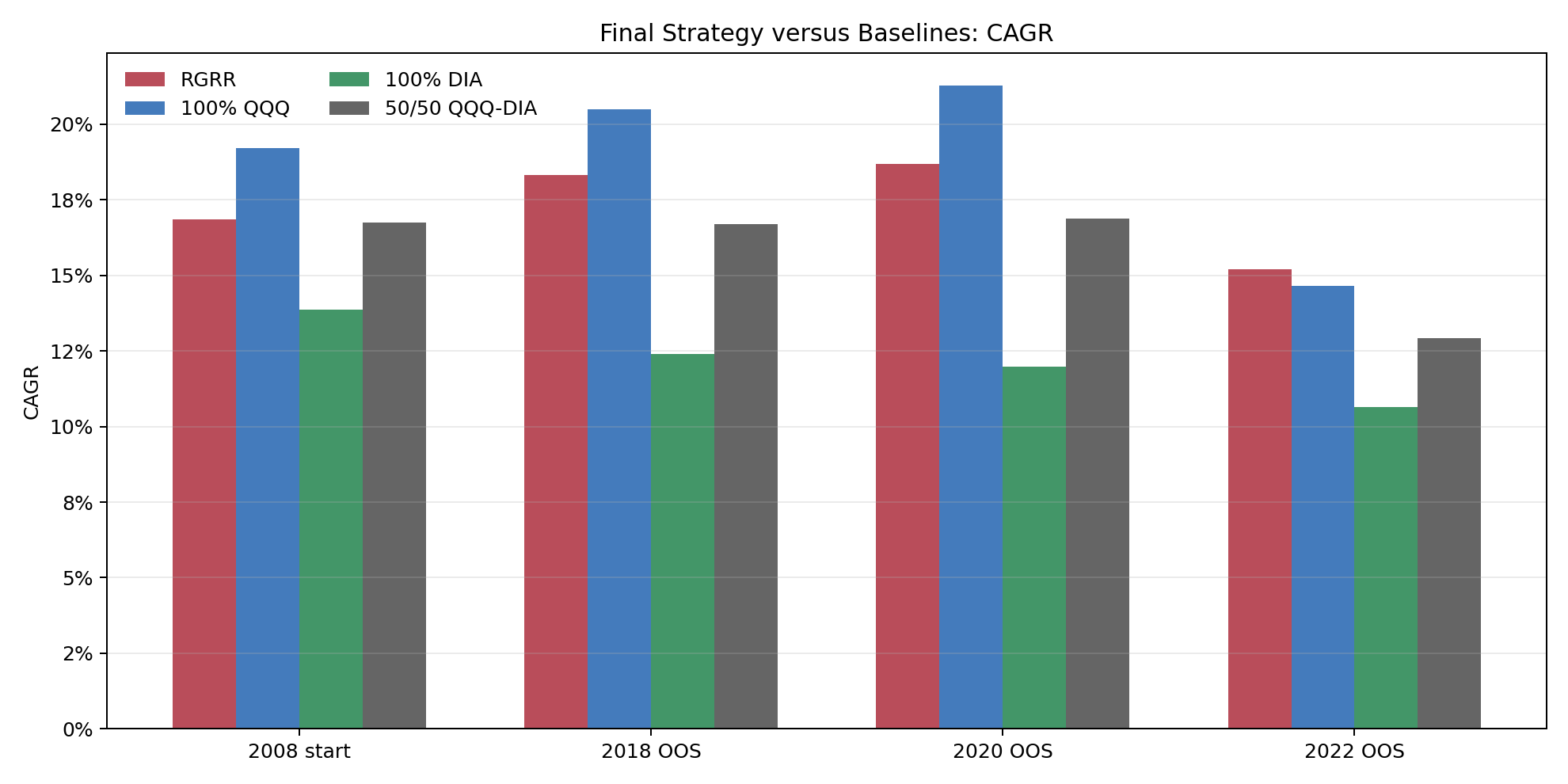}
\caption{CAGR by OOS period.}
\end{subfigure}
\vspace{0.6em}
\begin{subfigure}{0.98\textwidth}
\includegraphics[width=\textwidth]{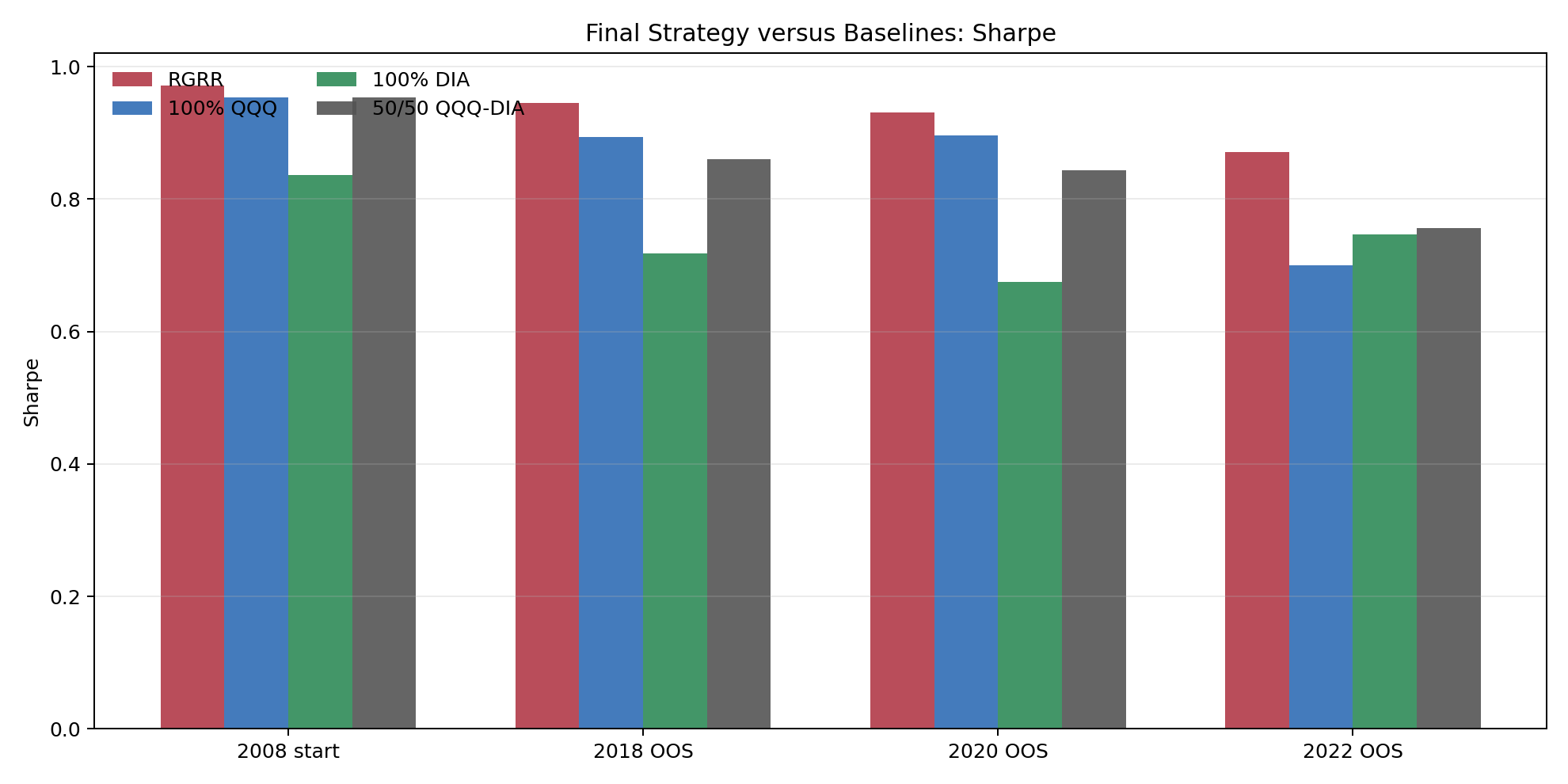}
\caption{Sharpe by OOS period.}
\end{subfigure}
\caption{RGRR versus Baselines across OOS Starts}
\label{fig:final-bars}
\end{figure}

The bar charts summarize the main conclusion. RGRR does not always beat QQQ on CAGR, but it improves Sharpe more consistently. This distinction is essential for interpreting the method. RGRR is not a proof that QQQ leadership should always be faded; it is evidence that the screened relative allocation score improves risk-adjusted exposure in the tested windows.

\section{Rolling Start-Date Sensitivity}

Table~\ref{tab:final-vs-baselines} reports the aligned baseline comparison for each tested interval. For the 2008 requested start, the actual rolling strategy start is 2009-07-27 because the rolling selector needs a 756-day training window and valid standardized features.

\begin{table}[H]
\centering
\caption{RGRR versus Baselines by Evaluation Window}
\label{tab:final-vs-baselines}
\begingroup
\tiny
\setlength{\tabcolsep}{3pt}
\resizebox{\textwidth}{!}{%
\begin{tabular}{llrrrrrr}
\toprule
Period & Portfolio & CAGR & Sharpe & Max DD & Ann. Turnover & Avg. QQQ & Final Wealth\\
\midrule
2008 start & RGRR & \pct{16.85} & 0.97 & \pct{-29.25} & \pct{505.89} & \pct{48.83} & 13.73\\
2008 start & 100\% QQQ & \pct{19.22} & 0.95 & \pct{-35.12} & \pct{0.00} & \pct{100.00} & 19.25\\
2008 start & 100\% DIA & \pct{13.88} & 0.84 & \pct{-36.71} & \pct{0.00} & \pct{0.00} & 8.90\\
2008 start & 50/50 QQQ--DIA & \pct{16.75} & 0.95 & \pct{-32.28} & \pct{0.00} & \pct{50.00} & 13.55\\
2018 OOS & RGRR & \pct{18.33} & 0.94 & \pct{-29.41} & \pct{444.60} & \pct{49.38} & 3.84\\
2018 OOS & 100\% QQQ & \pct{20.50} & 0.89 & \pct{-35.12} & \pct{0.00} & \pct{100.00} & 4.44\\
2018 OOS & 100\% DIA & \pct{12.41} & 0.72 & \pct{-36.71} & \pct{0.00} & \pct{0.00} & 2.55\\
2018 OOS & 50/50 QQQ--DIA & \pct{16.69} & 0.86 & \pct{-32.28} & \pct{0.00} & \pct{50.00} & 3.43\\
2020 OOS & RGRR & \pct{18.69} & 0.93 & \pct{-29.41} & \pct{408.16} & \pct{50.37} & 3.03\\
2020 OOS & 100\% QQQ & \pct{21.29} & 0.90 & \pct{-35.12} & \pct{0.00} & \pct{100.00} & 3.49\\
2020 OOS & 100\% DIA & \pct{11.98} & 0.67 & \pct{-36.71} & \pct{0.00} & \pct{0.00} & 2.08\\
2020 OOS & 50/50 QQQ--DIA & \pct{16.88} & 0.84 & \pct{-32.28} & \pct{0.00} & \pct{50.00} & 2.75\\
2022 OOS & RGRR & \pct{15.19} & 0.87 & \pct{-23.62} & \pct{354.33} & \pct{54.03} & 1.88\\
2022 OOS & 100\% QQQ & \pct{14.65} & 0.70 & \pct{-34.83} & \pct{0.00} & \pct{100.00} & 1.84\\
2022 OOS & 100\% DIA & \pct{10.64} & 0.75 & \pct{-20.76} & \pct{0.00} & \pct{0.00} & 1.57\\
2022 OOS & 50/50 QQQ--DIA & \pct{12.93} & 0.76 & \pct{-26.80} & \pct{0.00} & \pct{50.00} & 1.72\\
\bottomrule
\end{tabular}
}
\endgroup
\end{table}

The most important comparison is not a single row but the pattern across rows. Against 100\% QQQ, RGRR gives up CAGR in the long 2008, 2018, and 2020 windows, but improves Sharpe and drawdown. Against 50/50 QQQ--DIA, RGRR improves CAGR, Sharpe, and drawdown in every tested interval. Against 100\% DIA, it improves return and Sharpe but not always drawdown, especially in the 2022 window where DIA is naturally defensive.

\begin{table}[H]
\centering
\caption{RGRR Delta versus Aligned Baselines}
\label{tab:baseline-deltas}
\scriptsize
\begin{tabular}{llrrrr}
\toprule
Period & Baseline & $\Delta$ CAGR & $\Delta$ Sharpe & DD Improvement & $\Delta$ Final Wealth\\
\midrule
2008 start & 100\% QQQ & \pct{-2.37} & 0.02 & \pct{5.87} & -5.52\\
2008 start & 100\% DIA & \pct{2.97} & 0.13 & \pct{7.46} & 4.83\\
2008 start & 50/50 QQQ--DIA & \pct{0.09} & 0.02 & \pct{3.03} & 0.18\\
2018 OOS & 100\% QQQ & \pct{-2.17} & 0.05 & \pct{5.71} & -0.60\\
2018 OOS & 100\% DIA & \pct{5.92} & 0.23 & \pct{7.30} & 1.29\\
2018 OOS & 50/50 QQQ--DIA & \pct{1.64} & 0.09 & \pct{2.87} & 0.41\\
2020 OOS & 100\% QQQ & \pct{-2.60} & 0.04 & \pct{5.71} & -0.46\\
2020 OOS & 100\% DIA & \pct{6.70} & 0.26 & \pct{7.30} & 0.95\\
2020 OOS & 50/50 QQQ--DIA & \pct{1.80} & 0.09 & \pct{2.87} & 0.29\\
2022 OOS & 100\% QQQ & \pct{0.55} & 0.17 & \pct{11.21} & 0.04\\
2022 OOS & 100\% DIA & \pct{4.55} & 0.12 & \pct{-2.86} & 0.31\\
2022 OOS & 50/50 QQQ--DIA & \pct{2.26} & 0.11 & \pct{3.18} & 0.16\\
\bottomrule
\end{tabular}
\end{table}

The delta table makes the core empirical claim sharper. RGRR's Sharpe delta versus QQQ is positive in every row, while its CAGR delta versus QQQ is positive only in the 2022 OOS row. Against 50/50, all three dimensions are positive in every tested interval. This is why the method should be evaluated as a conditional relative-rotation rule, not as a permanent QQQ substitute.

\subsection{Window-Level Interpretation}

The rolling results should be read together with the market environment of each requested start. Table~\ref{tab:window-interpretation} summarizes the economic interpretation of the main OOS windows. This table is deliberately qualitative, but each statement is anchored in the quantitative results above.

\begin{table}[H]
\centering
\caption{Window-Level Diagnostic Interpretation}
\label{tab:window-interpretation}
\scriptsize
\resizebox{\textwidth}{!}{%
\begin{tabular}{p{2.4cm}p{4.3cm}p{4.5cm}p{3.5cm}}
\toprule
Window & Baseline Environment & RGRR Behavior & Interpretation\\
\midrule
2018 OOS & QQQ has the highest terminal wealth, but the window includes the late-2018 drawdown, the 2020 crash, and the 2022 rate shock. & RGRR tilts around neutral, earns lower CAGR than QQQ, but improves Sharpe and drawdown while beating 50/50 on all reported dimensions. & The strategy works as a risk-adjusted relative allocator, not as an unconditional QQQ return enhancer.\\
2020 OOS & The sample begins just before a sharp crash and an exceptionally strong technology-led rebound. & RGRR avoids some risk but does not hold full QQQ through the strongest rebound, so CAGR trails QQQ while Sharpe improves. & This is the main negative test against an overstated timing claim. The method reduces risk but misses part of the high-growth recovery.\\
2022 OOS & Rising-rate stress and style rotation make QQQ a much weaker benchmark than in the earlier bull-market windows. & RGRR improves CAGR, Sharpe, and maximum drawdown versus QQQ, while also beating 50/50 on return and Sharpe. & This is the clearest window for the relief-gated relative-rotation mechanism.\\
2008 start & The long robustness window contains both crisis recovery and a long technology-led bull market. & RGRR ends near 50/50 terminal wealth and below QQQ, but with slightly higher Sharpe than QQQ and 50/50. & Long-run results support a bounded Sharpe claim rather than a terminal-wealth dominance claim.\\
\bottomrule
\end{tabular}
}
\end{table}

This window-level reading also explains why the paper keeps 100\% QQQ as a hard benchmark. If the strategy only beat DIA or 50/50, the result would be less informative because QQQ is the stronger unconditional asset in this sample. The final evidence is more nuanced: RGRR can improve the risk-adjusted path relative to QQQ, but it should not be expected to compound faster than QQQ during every growth-led market.

\begin{table}[H]
\centering
\caption{Rolling Selection Stability by Requested Start}
\label{tab:selection-stability}
\scriptsize
\resizebox{\textwidth}{!}{%
\begin{tabular}{lrrrrrr}
\toprule
Period & Blocks & Unique Configs & Top Share & Median Train CAGR & Median Train Sharpe & Median Train Turnover\\
\midrule
2008 start & 68 & 17 & \pct{25.00} & \pct{15.44} & 0.91 & \pct{531.19}\\
2018 OOS & 32 & 12 & \pct{25.00} & \pct{18.11} & 0.97 & \pct{466.61}\\
2020 OOS & 26 & 8 & \pct{30.77} & \pct{17.45} & 0.83 & \pct{513.34}\\
2022 OOS & 18 & 6 & \pct{27.78} & \pct{13.63} & 0.80 & \pct{452.06}\\
\bottomrule
\end{tabular}
}
\end{table}

The selection-stability table is important because the rolling strategy is not a fixed-parameter holdout. In each 63-day block, the signal universe and mapping remain fixed, but the lambda combination can change. Table~\ref{tab:selection-stability} shows that the selector uses several configurations across windows rather than collapsing to one static configuration. This is expected for a rolling relative-allocation rule: the admitted economic vocabulary is stable, while the emphasis on main, second-order, and conditional relief terms adapts to recent training evidence.

\section{Expanding Start-Date Sensitivity}

The final formal validation in this paper is rolling, because the strategy is meant to adapt to changing market structure. Nevertheless, the prior cash-overlay study reports expanding diagnostics \cite{xiong2026cashoverlay}, so the same diagnostic is included here for format and robustness. In the expanding version, the signal universe, fixed mapping, cost convention, and lambda grid are identical to RGRR, but each training set uses all available past data before the test block rather than the most recent 756 trading days. This section is therefore not a competing final strategy. It is a check of whether the fixed signal universe is fragile to the training-window convention.

\begin{table}[H]
\centering
\caption{Expanding Diagnostic for RGRR Signal Universe}
\label{tab:expanding-diagnostic}
\scriptsize
\begin{tabular}{llrrrrr}
\toprule
Period & Actual Start & CAGR & Sharpe & Max DD & Ann. Turnover & Avg. QQQ\\
\midrule
2008 start & 2009-07-27 & \pct{16.84} & 0.97 & \pct{-29.10} & \pct{478.89} & \pct{47.95}\\
2018 OOS & 2018-06-28 & \pct{18.15} & 0.94 & \pct{-29.10} & \pct{380.75} & \pct{48.39}\\
2020 OOS & 2020-01-02 & \pct{18.42} & 0.92 & \pct{-29.10} & \pct{332.20} & \pct{48.70}\\
2022 OOS & 2022-01-03 & \pct{15.19} & 0.88 & \pct{-23.60} & \pct{311.43} & \pct{52.51}\\
\bottomrule
\end{tabular}
\end{table}

The expanding diagnostic is close to the rolling result but has lower turnover in the later OOS windows. This is expected: expanding training changes more slowly because older observations remain in the training set. The result supports the fixed signal universe but does not overturn the final choice of rolling validation. The final strategy remains rolling because the research question concerns adaptive relative allocation rather than a permanently frozen historical mapping.

\section{Economic Interpretation and Limitations}

The empirical result is narrower than the initial intuitive story. The data do not support a simple statement that DIA catches up whenever QQQ has substantially led and then stalls. RGRR says something more conditional: relative reversal has value when combined with rate relief and, in one retained third-order term, broad market drawdown. Some statistically strong credit and volatility interactions explain regimes but do not add incremental OOS Sharpe once the lower-order policy is present.

This distinction is useful. It means RGRR has partially falsified the naive version of the QQQ-lead/DIA-catch-up story. It has not falsified all DIA catch-up logic. Instead, it narrows that logic to rate-relief and relative-reversal environments. The retained third-order terms are therefore not arbitrary complexity; they identify the specific context in which the relative-reversal mechanism survives the incremental OOS test.

The main weakness is turnover. RGRR has annualized turnover between 354\% and 506\% depending on the start window. The 10bp cost assumption is included, but higher real trading costs, tax effects, bid-ask spreads, and slippage could reduce the economic value. RGRR should therefore be interpreted as a research prototype. A lower-turnover version or a trade gate is a natural next experiment.

The second limitation is multiple testing. The global HAC screens and incremental OOS filter reduce overfitting risk, but they do not eliminate it. The current paper does not yet report a White Reality Check, Hansen SPA test, block-bootstrap inference, or Deflated Sharpe Ratio for the entire search process. The proper conclusion is that the method is promising and internally disciplined, not that it is a final production anomaly claim.

\section{Conclusion}

This paper builds RGRR, a QQQ--DIA relative allocation strategy using the same research discipline as the prior growth--defensive timing and cash-overlay allocation studies \cite{xiong2026growthdefensive,xiong2026cashoverlay}. The method admits statistically screened main and second-order signals, then applies a stricter incremental OOS filter to third-order interactions. The final formal signal stack contains one main effect, nine second-order interactions, and two third-order interactions.

The results support a risk-adjusted allocation claim. RGRR improves Sharpe versus 100\% QQQ and 50/50 QQQ--DIA across the tested OOS windows. It improves CAGR versus 50/50 in every interval and versus QQQ only in the 2022 OOS interval. This is consistent with a relative timing rule that reduces QQQ exposure in selected regimes, rather than a rule that always beats a high-return QQQ benchmark.

The next research step is to test lower-turnover variants, add formal data-snooping corrections, and evaluate whether the same screened relative-allocation design generalizes to other growth-versus-value ETF pairs.

\appendix
\section{Screening Tables and Robustness Details}

\subsection{Screening Protocol}

The screening layer is diagnostic rather than a direct trading rule. Each candidate main effect or interaction is evaluated against future QQQ--DIA relative returns at 21, 63, and 126 trading-day horizons. The best horizon per candidate is retained when the absolute HAC $t$-statistic is at least 2.0. Because the future-return labels overlap at daily frequency, ordinary OLS standard errors are not used as the primary inference object. The reported $t$-statistics use Newey--West/HAC standard errors.

For a candidate variable $x_t$, the diagnostic regression is
\begin{equation}
R^{Q-D}_{t,t+h}=\alpha_h+\beta_hx_t+\epsilon_{t,h},\qquad h\in\{21,63,126\}.
\end{equation}
The coefficient sign determines the trading orientation:
\begin{equation}
o_j=
\begin{cases}
+1, & \hat{\beta}_j>0,\\
-1, & \hat{\beta}_j<0.
\end{cases}
\end{equation}
The oriented feature is $o_jx_{j,t}$. A positive oriented feature therefore always means a state associated with stronger future QQQ relative performance, even when the raw regression coefficient was negative.

The de-duplication step removes candidates whose absolute correlation with a previously admitted candidate exceeds 0.95. This is a pragmatic guard against adding many copies of the same state under different labels. It is not a full multiple-testing correction. The final paper therefore keeps the inference claim bounded: the statistical screen defines a stable signal vocabulary, and the OOS portfolio tests evaluate whether that vocabulary is useful as a trading rule.

\subsection{Lower-Order Signal Families}

The lower-order filter is the direct analogue of the component-construction sections in the cash-overlay study \cite{xiong2026cashoverlay}. In the present two-ETF relative allocation problem, there is no cash overlay and no crash-brake module. The equivalent object is a relative-allocation score built from a main effect and second-order interactions. Table~\ref{tab:lower-order-families} groups the admitted lower-order terms by economic role.

\begin{table}[H]
\centering
\caption{Lower-Order RGRR Signal Families}
\label{tab:lower-order-families}
\scriptsize
\begin{tabular}{p{3.0cm}p{5.5cm}p{5.5cm}}
\toprule
Family & Terms & Economic Interpretation\\
\midrule
Rate relief & \code{rate_relief} & Falling long rates are the only main effect that survives the global screen.\\
Rate plus calm/stress & \code{rate_relief * low_vix}; \code{rate_relief * spy_drawdown} & Rate relief has different implications depending on whether volatility is calm or the broad market is drawn down.\\
Relative reversal plus rate & \code{rel_reversal * rate_relief} & The simplest lower-order version of the DIA catch-up channel: QQQ relative weakness matters when rates are easing.\\
Relative reversal plus stress/credit & \code{rel_reversal * high_vix}; \code{rel_reversal * credit_relief} & Reversal pressure is more informative when volatility or credit conditions define the broader state.\\
Relative momentum plus credit & \code{rel_mom126 * credit_relief} & Prior QQQ relative momentum has a different implication when credit appetite improves.\\
Volatility and credit relief/stress & \code{vix_relief * credit_stress}; \code{credit_relief * credit_stress}; \code{rate_relief * credit_relief} & Relief and stress variables interact rather than acting as simple one-direction predictors.\\
\bottomrule
\end{tabular}
\end{table}

This family view is useful because several interactions have negative raw coefficients. After orientation, they can still contribute positively to the RGRR score. The economics should therefore be read through the oriented score, not by manually reading every raw sign in isolation.

\subsection{Global Interaction Screen before Incremental Filtering}

Table~\ref{tab:global-interaction-screen} reports the complete globally screened interaction set before the additional third-order incremental OOS filter. This table is intentionally larger than the formal final signal set. It separates two concepts that are easy to confuse: statistical admission into the diagnostic interaction set and portfolio admission into the final trading strategy. The nine second-order terms enter RGRR after this screen. The third-order terms must pass the extra incremental filter reported later.

\begingroup
\tiny
\setlength{\tabcolsep}{3pt}
\begin{longtable}{rp{8.4cm}rrr}
\caption{Complete Globally Screened Interaction Set}
\label{tab:global-interaction-screen}\\
\toprule
Order & Terms & Horizon & HAC $t$ & Orientation\\
\midrule
\endfirsthead
\toprule
Order & Terms & Horizon & HAC $t$ & Orientation\\
\midrule
\endhead
2 & vix\_relief $\times$ credit\_stress & 126 & -3.62 & -1 \\
2 & rel\_mom126 $\times$ credit\_relief & 126 & -2.79 & -1 \\
2 & rate\_relief $\times$ low\_vix & 21 & -2.75 & -1 \\
2 & rate\_relief $\times$ credit\_relief & 126 & -2.69 & -1 \\
2 & rel\_reversal $\times$ high\_vix & 126 & -2.50 & -1 \\
2 & rate\_relief $\times$ spy\_drawdown & 126 & 2.48 & 1 \\
2 & rel\_reversal $\times$ rate\_relief & 63 & 2.47 & 1 \\
2 & credit\_relief $\times$ credit\_stress & 21 & -2.31 & -1 \\
2 & rel\_reversal $\times$ credit\_relief & 126 & 2.19 & 1 \\
3 & vix\_relief $\times$ credit\_relief $\times$ credit\_stress & 126 & 7.16 & 1 \\
3 & spy\_drawdown $\times$ vix\_relief $\times$ credit\_relief & 126 & 7.11 & 1 \\
3 & rate\_relief $\times$ vix\_relief $\times$ credit\_relief & 126 & 6.55 & 1 \\
3 & rate\_relief $\times$ credit\_relief $\times$ credit\_stress & 126 & -5.50 & -1 \\
3 & spy\_drawdown $\times$ credit\_relief $\times$ credit\_stress & 126 & -4.88 & -1 \\
3 & rel\_mom126 $\times$ spy\_drawdown $\times$ credit\_relief & 126 & -4.63 & -1 \\
3 & spy\_drawdown $\times$ vix\_relief $\times$ low\_vix & 126 & 4.61 & 1 \\
3 & spy\_drawdown $\times$ high\_vix $\times$ credit\_stress & 126 & 4.55 & 1 \\
3 & rel\_mom126 $\times$ rate\_relief $\times$ vix\_relief & 126 & -4.26 & -1 \\
3 & rate\_relief $\times$ high\_vix $\times$ credit\_relief & 126 & -4.17 & -1 \\
3 & rel\_reversal $\times$ rate\_relief $\times$ spy\_drawdown & 63 & 4.09 & 1 \\
3 & rel\_mom126 $\times$ rate\_relief $\times$ credit\_relief & 126 & -3.87 & -1 \\
3 & rate\_relief $\times$ spy\_drawdown $\times$ credit\_stress & 126 & 3.75 & 1 \\
3 & rel\_reversal $\times$ spy\_drawdown $\times$ vix\_relief & 21 & 3.67 & 1 \\
3 & rel\_mom126 $\times$ vix\_relief $\times$ credit\_relief & 126 & 3.52 & 1 \\
3 & high\_vix $\times$ vix\_relief $\times$ credit\_relief & 126 & 3.38 & 1 \\
3 & rel\_mom126 $\times$ credit\_relief $\times$ credit\_stress & 63 & -3.06 & -1 \\
3 & rel\_mom126 $\times$ rel\_reversal $\times$ rate\_relief & 126 & -2.78 & -1 \\
3 & rate\_relief $\times$ high\_vix $\times$ low\_vix & 126 & -2.77 & -1 \\
3 & rate\_relief $\times$ vix\_relief $\times$ low\_vix & 63 & 2.75 & 1 \\
3 & rel\_mom126 $\times$ spy\_drawdown $\times$ high\_vix & 126 & 2.55 & 1 \\
3 & rel\_mom126 $\times$ high\_vix $\times$ credit\_stress & 126 & 2.31 & 1 \\
3 & rel\_reversal $\times$ spy\_drawdown $\times$ credit\_relief & 126 & 2.29 & 1 \\
3 & rate\_relief $\times$ spy\_drawdown $\times$ vix\_relief & 21 & -2.16 & -1 \\
3 & rate\_relief $\times$ spy\_drawdown $\times$ low\_vix & 126 & -2.13 & -1 \\
3 & rel\_reversal $\times$ rate\_relief $\times$ vix\_relief & 126 & 2.00 & 1 \\
\bottomrule
\end{longtable}
\endgroup

\subsection{Third-Order Candidate Families}

Third-order interactions are formed from the same state vocabulary but are treated more conservatively. The candidate families are organized by the economic groups appearing in their terms: relative state, rate state, volatility state, credit state, and broad-market drawdown. Table~\ref{tab:ix3-family-examples} reports representative families and the final admission outcome.

\begin{table}[H]
\centering
\caption{Third-Order Candidate Families and Admission Outcome}
\label{tab:ix3-family-examples}
\scriptsize
\begin{tabular}{p{3.2cm}p{5.3cm}p{5.4cm}}
\toprule
Family & Representative Terms & Outcome\\
\midrule
Rate + relative & \code{rel_mom126 * rel_reversal * rate_relief} & Retained. It has positive mean OOS Sharpe delta and is positive in all three screen periods.\\
Market drawdown + rate + relative & \code{rel_reversal * rate_relief * spy_drawdown} & Retained. It passes the two-of-three positive-period rule and adds a small complementary effect.\\
Credit + volatility & \code{vix_relief * credit_relief * credit_stress} & Rejected as trading signal despite strong HAC statistics because incremental OOS Sharpe is negative.\\
Credit + market drawdown + volatility & \code{spy_drawdown * vix_relief * credit_relief} & Rejected. The state is interpretable but does not improve the lower-order portfolio.\\
Rate + volatility & \code{rate_relief * high_vix * low_vix}; \code{rate_relief * vix_relief * low_vix} & Rejected. The family is too fragile after the lower-order rate and volatility terms are already present.\\
Credit + relative & \code{rel_mom126 * credit_relief * credit_stress} & Rejected. Credit-relative interactions remain diagnostic but not tradable in the final stack.\\
\bottomrule
\end{tabular}
\end{table}

The outcome explains why the final method name should not be a technical name such as \code{filtered_ix3_top2}. The economic method is not ``use two third-order terms.'' It is to rotate between QQQ and DIA only when relative states are gated by macro relief conditions that survive incremental OOS validation.

\subsection{From Diagnostic Scores to Tradable Weights}

The score-to-weight transformation is identical across rolling and expanding diagnostics. First, admitted features are oriented and averaged within their signal group. Second, each group score is expanding-standardized. Third, the rolling selector chooses lambdas on the admitted groups. Fourth, the combined score is mapped into a lagged and smoothed QQQ weight.

\begin{align}
S^{RGRR}_t &= \lambda_m S^{main}_t+\lambda_2 S^{ix2}_t+\lambda_3 S^{ix3}_t,\\
\tilde{w}^Q_t &= 0.5+0.50\tanh\left(Z(S^{RGRR}_t)/0.75\right),\\
w^Q_t &= 0.95w^Q_{t-1}+0.05\tilde{w}^Q_{t-1}.
\end{align}
The realized net return is
\begin{equation}
r^P_t=w^Q_tr^Q_t+(1-w^Q_t)r^D_t-0.0010\cdot 2|w^Q_t-w^Q_{t-1}|.
\end{equation}
The one-day lag means that today's close-to-close return uses a weight determined by information available before today's return is realized. The smoothing parameter keeps the strategy from jumping from all DIA to all QQQ in one day, but it also contributes to lag and turnover. This is why the paper identifies turnover as the main implementation limitation.

\subsection{Third-Order Candidate Screen}

Table~\ref{tab:ix3-candidate-screen} reports the leading third-order candidates by mean rolling OOS Sharpe delta versus the main+ix2 policy. Only the first two candidates pass the final incremental rule. The table demonstrates why statistical significance alone is insufficient for third-order terms: several interactions have large absolute HAC $t$-statistics but negative incremental OOS contribution.

\begin{table}[H]
\centering
\caption{Top Third-Order Candidate Incremental OOS Screen}
\label{tab:ix3-candidate-screen}
\begingroup
\tiny
\setlength{\tabcolsep}{2pt}
\resizebox{\textwidth}{!}{%
\begin{tabular}{p{5.7cm}p{3.0cm}rrrr}
\toprule
Interaction & Family & Mean $\Delta$ Sharpe & Positive Periods & Mean $\Delta$ CAGR & Pass\\
\midrule
\code{rel_mom126 * rel_reversal * rate_relief} & rate + relative & 0.019 & 3 & \pct{0.50} & Yes\\
\code{rel_reversal * rate_relief * spy_drawdown} & market dd + rate + relative & 0.000 & 2 & \pct{0.14} & Yes\\
\code{rel_reversal * spy_drawdown * credit_relief} & credit + market dd + relative & -0.002 & 1 & \pct{-0.06} & No\\
\code{rel_reversal * spy_drawdown * vix_relief} & market dd + relative + vol & -0.002 & 1 & \pct{-0.13} & No\\
\code{rate_relief * spy_drawdown * vix_relief} & market dd + rate + vol & -0.004 & 1 & \pct{-0.23} & No\\
\code{rate_relief * high_vix * credit_relief} & credit + rate + vol & -0.005 & 0 & \pct{-0.21} & No\\
\code{rel_mom126 * credit_relief * credit_stress} & credit + relative & -0.006 & 1 & \pct{-0.21} & No\\
\code{rate_relief * spy_drawdown * credit_stress} & credit + market dd + rate & -0.007 & 0 & \pct{-0.24} & No\\
\code{rate_relief * vix_relief * low_vix} & rate + vol & -0.008 & 0 & \pct{-0.24} & No\\
\code{spy_drawdown * vix_relief * low_vix} & market dd + vol & -0.008 & 1 & \pct{-0.28} & No\\
\code{spy_drawdown * vix_relief * credit_relief} & credit + market dd + vol & -0.009 & 0 & \pct{-0.31} & No\\
\code{rel_mom126 * spy_drawdown * credit_relief} & credit + market dd + relative & -0.009 & 0 & \pct{-0.33} & No\\
\bottomrule
\end{tabular}
}
\endgroup
\end{table}

\begingroup
\tiny
\setlength{\tabcolsep}{2pt}
\begin{longtable}{p{5.2cm}p{3.0cm}rrrrr}
\caption{Full Third-Order Incremental OOS Candidate Table}
\label{tab:ix3-full-candidates}\\
\toprule
Terms & Family & Mean $\Delta$ Sharpe & Positive Periods & Mean $\Delta$ CAGR & HAC $|t|$ & Pass\\
\midrule
\endfirsthead
\toprule
Terms & Family & Mean $\Delta$ Sharpe & Positive Periods & Mean $\Delta$ CAGR & HAC $|t|$ & Pass\\
\midrule
\endhead
rel\_mom126 $\times$ rel\_reversal $\times$ rate\_relief & rate + relative & 0.019 & 3 & \pct{0.50} & 2.78 & Yes \\
rel\_reversal $\times$ rate\_relief $\times$ spy\_drawdown & market\_dd + rate + relative & 0.000 & 2 & \pct{0.14} & 4.09 & Yes \\
rel\_reversal $\times$ spy\_drawdown $\times$ credit\_relief & credit + market\_dd + relative & -0.002 & 1 & \pct{-0.06} & 2.29 & No \\
rel\_reversal $\times$ spy\_drawdown $\times$ vix\_relief & market\_dd + relative + vol & -0.002 & 1 & \pct{-0.13} & 3.67 & No \\
rate\_relief $\times$ spy\_drawdown $\times$ vix\_relief & market\_dd + rate + vol & -0.004 & 1 & \pct{-0.23} & 2.16 & No \\
rate\_relief $\times$ high\_vix $\times$ credit\_relief & credit + rate + vol & -0.005 & 0 & \pct{-0.21} & 4.17 & No \\
rel\_mom126 $\times$ credit\_relief $\times$ credit\_stress & credit + relative & -0.006 & 1 & \pct{-0.21} & 3.06 & No \\
rate\_relief $\times$ spy\_drawdown $\times$ credit\_stress & credit + market\_dd + rate & -0.007 & 0 & \pct{-0.24} & 3.75 & No \\
rate\_relief $\times$ vix\_relief $\times$ low\_vix & rate + vol & -0.008 & 0 & \pct{-0.24} & 2.75 & No \\
spy\_drawdown $\times$ vix\_relief $\times$ low\_vix & market\_dd + vol & -0.008 & 1 & \pct{-0.28} & 4.61 & No \\
spy\_drawdown $\times$ vix\_relief $\times$ credit\_relief & credit + market\_dd + vol & -0.009 & 0 & \pct{-0.31} & 7.11 & No \\
rel\_mom126 $\times$ spy\_drawdown $\times$ credit\_relief & credit + market\_dd + relative & -0.009 & 0 & \pct{-0.33} & 4.63 & No \\
spy\_drawdown $\times$ high\_vix $\times$ credit\_stress & credit + market\_dd + vol & -0.009 & 0 & \pct{-0.37} & 4.55 & No \\
rate\_relief $\times$ credit\_relief $\times$ credit\_stress & credit + rate & -0.010 & 0 & \pct{-0.30} & 5.50 & No \\
high\_vix $\times$ vix\_relief $\times$ credit\_relief & credit + vol & -0.011 & 0 & \pct{-0.35} & 3.38 & No \\
rel\_mom126 $\times$ high\_vix $\times$ credit\_stress & credit + relative + vol & -0.011 & 0 & \pct{-0.29} & 2.31 & No \\
rate\_relief $\times$ spy\_drawdown $\times$ low\_vix & market\_dd + rate + vol & -0.011 & 0 & \pct{-0.30} & 2.13 & No \\
rel\_mom126 $\times$ vix\_relief $\times$ credit\_relief & credit + relative + vol & -0.013 & 0 & \pct{-0.32} & 3.52 & No \\
rate\_relief $\times$ vix\_relief $\times$ credit\_relief & credit + rate + vol & -0.013 & 0 & \pct{-0.36} & 6.55 & No \\
vix\_relief $\times$ credit\_relief $\times$ credit\_stress & credit + vol & -0.015 & 0 & \pct{-0.39} & 7.16 & No \\
spy\_drawdown $\times$ credit\_relief $\times$ credit\_stress & credit + market\_dd & -0.022 & 0 & \pct{-0.60} & 4.88 & No \\
rel\_mom126 $\times$ rate\_relief $\times$ credit\_relief & credit + rate + relative & -0.023 & 0 & \pct{-0.54} & 3.87 & No \\
rate\_relief $\times$ high\_vix $\times$ low\_vix & rate + vol & -0.024 & 0 & \pct{-0.51} & 2.77 & No \\
rel\_mom126 $\times$ spy\_drawdown $\times$ high\_vix & market\_dd + relative + vol & -0.031 & 0 & \pct{-0.62} & 2.55 & No \\
rel\_mom126 $\times$ rate\_relief $\times$ vix\_relief & rate + relative + vol & -0.032 & 0 & \pct{-0.69} & 4.26 & No \\
rel\_reversal $\times$ rate\_relief $\times$ vix\_relief & rate + relative + vol & -0.043 & 0 & \pct{-0.72} & 2.00 & No \\
\bottomrule
\end{longtable}
\endgroup

Table~\ref{tab:ix3-full-candidates} is the strongest evidence against admitting all statistically screened third-order terms. Many of the largest absolute HAC statistics are associated with negative incremental OOS Sharpe. The final method therefore treats statistical significance as a necessary diagnostic screen, not as a sufficient trading admission rule.

\subsection{Base Main+IX2 versus Filtered IX3}

The incremental filter is measured against the rolling \code{base_main_ix2} policy. Table~\ref{tab:ix3-set-summary} shows that the two-term filtered IX3 set improves CAGR and Sharpe versus the base in the 2018, 2020, and 2022 OOS windows. The long 2008-start row is kept as a robustness row rather than as a screen row. The phrase ``filtered IX3'' here is a reproducibility label for the experiment; the final method name remains RGRR.

\begin{table}[H]
\centering
\caption{Filtered IX3 Set versus Main+IX2 Base}
\label{tab:ix3-set-summary}
\scriptsize
\begin{tabular}{lrrrr}
\toprule
Period & $\Delta$ CAGR & $\Delta$ Sharpe & $\Delta$ Max DD & $\Delta$ Turnover\\
\midrule
2008 start & \pct{0.25} & 0.005 & \pct{-0.01} & \pct{5.26}\\
2018 OOS & \pct{0.68} & 0.018 & \pct{0.19} & \pct{-4.51}\\
2020 OOS & \pct{0.79} & 0.023 & \pct{0.19} & \pct{-17.14}\\
2022 OOS & \pct{1.00} & 0.028 & \pct{0.16} & \pct{-34.17}\\
\bottomrule
\end{tabular}
\end{table}

\subsection{Signal-Group Ablation}

The final strategy uses main, second-order, and retained third-order groups together. To check whether one group alone explains the result, Table~\ref{tab:signal-group-ablation} reports rolling OOS ablations on the 2018, 2020, and 2022 windows. Each row uses the same fixed mapping $(M,\tau,\eta)=(0.50,0.75,0.05)$ and the same 10bp cost convention, but restricts the signal universe to the listed group.

\begin{table}[H]
\centering
\caption{Rolling Signal-Group Ablation}
\label{tab:signal-group-ablation}
\begingroup
\tiny
\setlength{\tabcolsep}{2pt}
\resizebox{\textwidth}{!}{%
\begin{tabular}{llrrrrrr}
\toprule
Period & Signal Group & CAGR & Sharpe & Max DD & Ann. Turnover & Avg. QQQ & $\Delta$ Sharpe vs QQQ\\
\midrule
2018 OOS & Main only & \pct{17.94} & 0.92 & \pct{-28.86} & \pct{626.03} & \pct{50.43} & 0.03\\
2018 OOS & IX2 only & \pct{17.23} & 0.91 & \pct{-30.08} & \pct{301.66} & \pct{40.88} & 0.02\\
2018 OOS & IX3 only & \pct{17.30} & 0.90 & \pct{-29.30} & \pct{114.91} & \pct{49.64} & 0.00\\
2018 OOS & Main+IX2 & \pct{17.44} & 0.92 & \pct{-29.61} & \pct{476.08} & \pct{44.15} & 0.02\\
2018 OOS & All screened groups & \pct{17.32} & 0.91 & \pct{-29.52} & \pct{321.00} & \pct{43.60} & 0.02\\
2020 OOS & Main only & \pct{18.14} & 0.91 & \pct{-28.86} & \pct{626.65} & \pct{47.64} & 0.01\\
2020 OOS & IX2 only & \pct{17.47} & 0.89 & \pct{-30.08} & \pct{260.87} & \pct{42.39} & -0.00\\
2020 OOS & IX3 only & \pct{17.62} & 0.90 & \pct{-29.30} & \pct{69.42} & \pct{44.24} & 0.00\\
2020 OOS & Main+IX2 & \pct{17.70} & 0.90 & \pct{-29.61} & \pct{452.97} & \pct{44.59} & 0.00\\
2020 OOS & All screened groups & \pct{17.42} & 0.89 & \pct{-29.43} & \pct{267.18} & \pct{43.35} & -0.00\\
2022 OOS & Main only & \pct{13.79} & 0.80 & \pct{-23.71} & \pct{669.53} & \pct{48.18} & 0.10\\
2022 OOS & IX2 only & \pct{14.58} & 0.87 & \pct{-24.45} & \pct{258.90} & \pct{44.24} & 0.17\\
2022 OOS & IX3 only & \pct{13.00} & 0.78 & \pct{-25.69} & \pct{52.66} & \pct{42.68} & 0.08\\
2022 OOS & Main+IX2 & \pct{13.99} & 0.83 & \pct{-23.78} & \pct{424.68} & \pct{45.59} & 0.13\\
2022 OOS & All screened groups & \pct{13.59} & 0.82 & \pct{-23.99} & \pct{257.71} & \pct{43.34} & 0.12\\
\bottomrule
\end{tabular}
}
\endgroup
\end{table}

The ablation table has two implications. First, no single group is a complete solution. The main effect is strong in risk-adjusted terms but has high turnover; the second-order layer is helpful in 2022 but weaker in earlier windows; the retained third-order layer is not attractive as a standalone strategy. Second, allowing every screened group into the score is also not optimal. The all-screened version is weaker than the final RGRR design because many statistically strong third-order terms are redundant or harmful once lower-order signals are present. This is why the final rule uses a filtered high-order admission step rather than a simple all-signal overlay.

\subsection{Weight Path Examples}

\begin{figure}[H]
\centering
\includegraphics[width=0.98\textwidth]{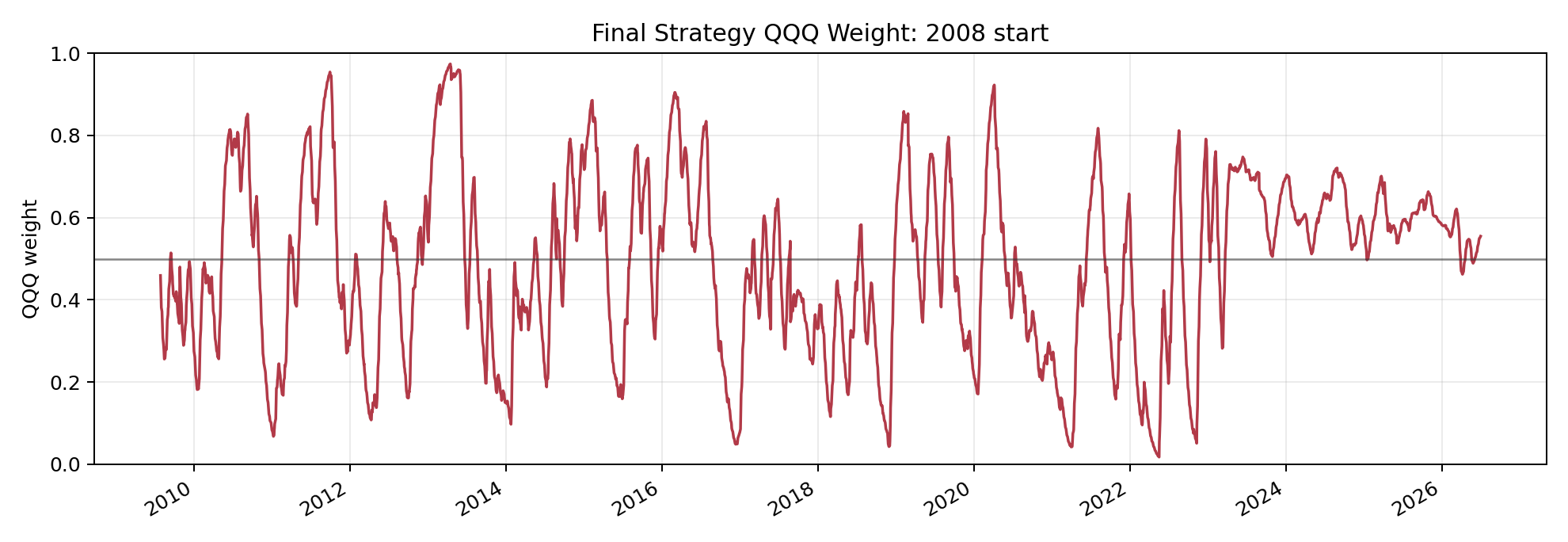}
\caption{RGRR QQQ Weight, Long 2008-Start Robustness Window}
\label{fig:weight-2008}
\end{figure}

Figure~\ref{fig:weight-2008} shows the full long-sample rolling weight path after the first valid 2008-start block. The series is centered near 50\%, but it is not static. Long periods of higher QQQ exposure correspond to conditions in which the oriented relief score favors QQQ relative return, while lower QQQ exposure corresponds to states in which DIA becomes the preferred relative sleeve.

\begin{figure}[H]
\centering
\includegraphics[width=0.98\textwidth]{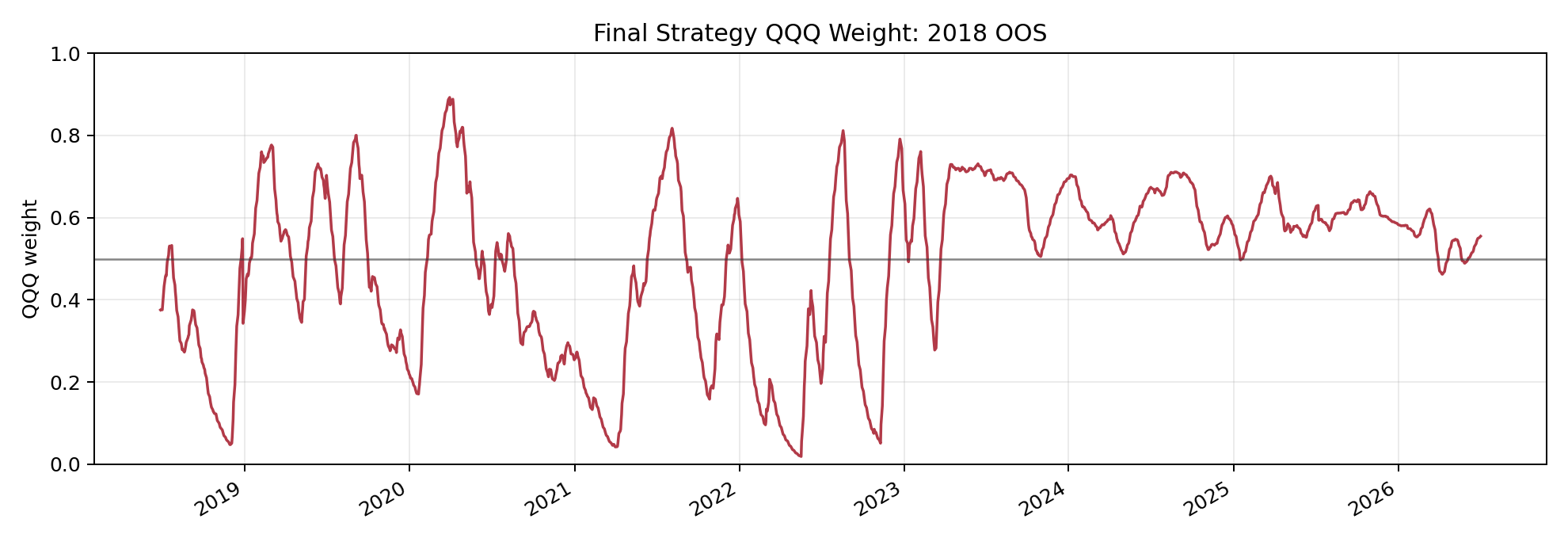}
\caption{RGRR QQQ Weight, 2018 OOS}
\label{fig:weight-2018}
\end{figure}

Figure~\ref{fig:weight-2018} shows RGRR's QQQ weight in the 2018 OOS window. The average QQQ weight is close to 50\%, but the allocation is not a static 50/50 portfolio. The strategy tilts around the neutral weight according to the fixed score mapping.

\begin{figure}[H]
\centering
\includegraphics[width=0.98\textwidth]{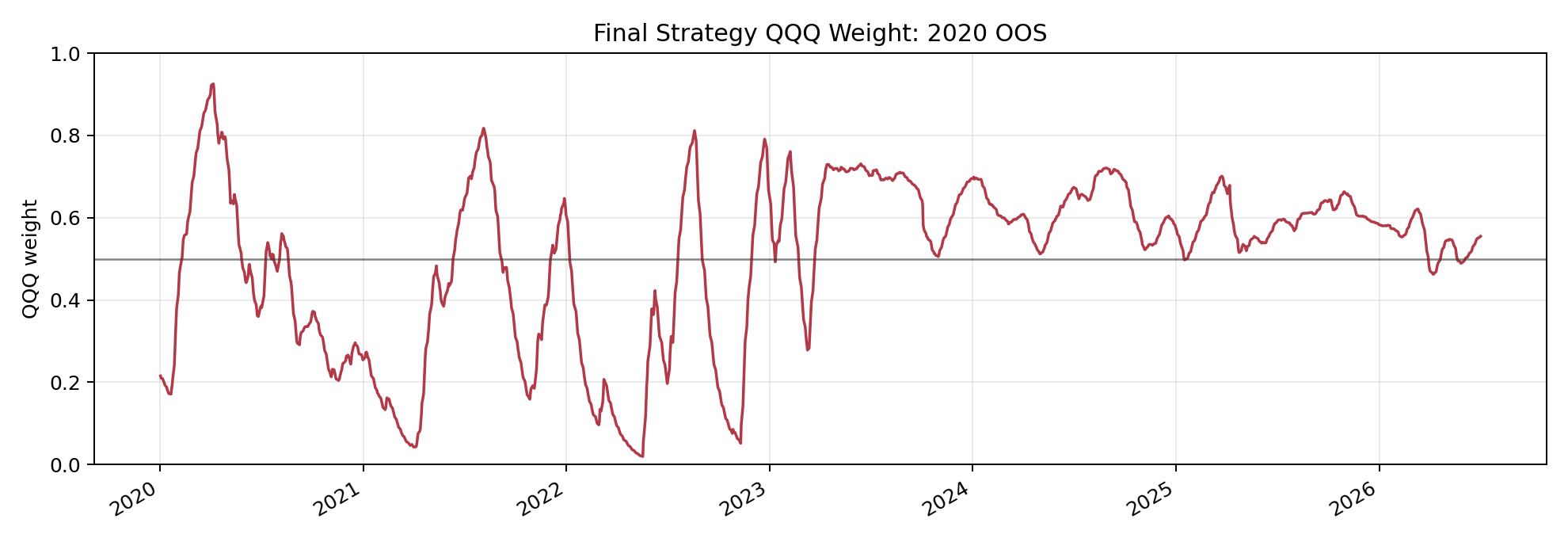}
\caption{RGRR QQQ Weight, 2020 OOS}
\label{fig:weight-2020}
\end{figure}

The 2020 weight path in Figure~\ref{fig:weight-2020} explains why RGRR does not beat QQQ on CAGR in that window. The strategy avoids some drawdown, but it also fails to hold full QQQ exposure through the strongest technology-led rebound. This is a useful negative diagnostic: the method is not simply buying every QQQ recovery.

\begin{figure}[H]
\centering
\includegraphics[width=0.98\textwidth]{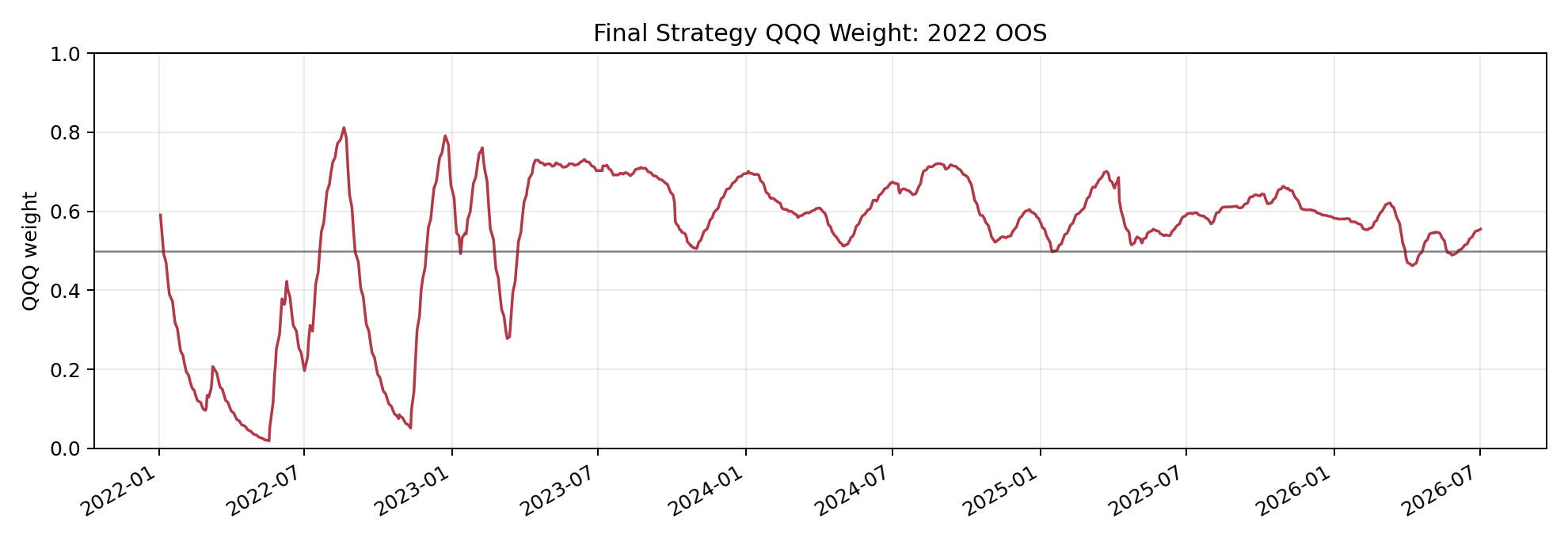}
\caption{RGRR QQQ Weight, 2022 OOS}
\label{fig:weight-2022}
\end{figure}

The 2022 path in Figure~\ref{fig:weight-2022} shows why RGRR can improve the 2022 window versus QQQ. The strategy spends much of the early stress period below full QQQ exposure and then gradually restores QQQ weight as the oriented relief score improves. This is not a binary regime classifier; it is a continuous score-to-weight policy with fixed smoothing.

\subsection{Complete Annual Matrix}

Table~\ref{tab:complete-annual-matrix} reports the full calendar-year matrix for the long 2008-start rolling path. The abbreviated annual table in the main text focuses on the recent period; this appendix table keeps the full post-warmup sample for auditability.

\begingroup
\tiny
\setlength{\tabcolsep}{3pt}
\begin{longtable}{llrrrr}
\caption{Complete Annual Matrix, 2008-Start Rolling Path}
\label{tab:complete-annual-matrix}\\
\toprule
Year & Portfolio & Return & Sharpe & Max DD & Avg. QQQ\\
\midrule
\endfirsthead
\toprule
Year & Portfolio & Return & Sharpe & Max DD & Avg. QQQ\\
\midrule
\endhead
2009 & RGRR & \pct{40.43} & 2.50 & \pct{-4.54} & \pct{38.39} \\
2009 & 100\% QQQ & \pct{48.69} & 2.50 & \pct{-5.53} & \pct{100.00} \\
2009 & 100\% DIA & \pct{36.16} & 2.39 & \pct{-3.77} & \pct{0.00} \\
2009 & 50/50 QQQ-DIA & \pct{42.41} & 2.53 & \pct{-4.60} & \pct{50.00} \\
2010 & RGRR & \pct{15.88} & 0.92 & \pct{-14.13} & \pct{50.68} \\
2010 & 100\% QQQ & \pct{17.02} & 0.90 & \pct{-15.29} & \pct{100.00} \\
2010 & 100\% DIA & \pct{11.46} & 0.75 & \pct{-13.17} & \pct{0.00} \\
2010 & 50/50 QQQ-DIA & \pct{14.30} & 0.85 & \pct{-14.17} & \pct{50.00} \\
2011 & RGRR & \pct{8.91} & 0.48 & \pct{-16.74} & \pct{54.57} \\
2011 & 100\% QQQ & \pct{-2.77} & 0.01 & \pct{-18.87} & \pct{100.00} \\
2011 & 100\% DIA & \pct{28.07} & 0.94 & \pct{-16.34} & \pct{0.00} \\
2011 & 50/50 QQQ-DIA & \pct{12.37} & 0.59 & \pct{-16.16} & \pct{50.00} \\
2012 & RGRR & \pct{11.65} & 0.93 & \pct{-9.62} & \pct{38.74} \\
2012 & 100\% QQQ & \pct{15.56} & 1.01 & \pct{-12.75} & \pct{100.00} \\
2012 & 100\% DIA & \pct{7.29} & 0.66 & \pct{-8.88} & \pct{0.00} \\
2012 & 50/50 QQQ-DIA & \pct{11.45} & 0.90 & \pct{-10.65} & \pct{50.00} \\
2013 & RGRR & \pct{29.26} & 2.32 & \pct{-5.64} & \pct{59.38} \\
2013 & 100\% QQQ & \pct{37.10} & 2.62 & \pct{-5.92} & \pct{100.00} \\
2013 & 100\% DIA & \pct{26.75} & 2.39 & \pct{-5.85} & \pct{0.00} \\
2013 & 50/50 QQQ-DIA & \pct{31.91} & 2.65 & \pct{-5.30} & \pct{50.00} \\
2014 & RGRR & \pct{11.99} & 1.02 & \pct{-7.75} & \pct{46.29} \\
2014 & 100\% QQQ & \pct{19.36} & 1.34 & \pct{-8.32} & \pct{100.00} \\
2014 & 100\% DIA & \pct{8.68} & 0.83 & \pct{-6.94} & \pct{0.00} \\
2014 & 50/50 QQQ-DIA & \pct{13.98} & 1.17 & \pct{-7.46} & \pct{50.00} \\
2015 & RGRR & \pct{4.96} & 0.38 & \pct{-13.87} & \pct{54.14} \\
2015 & 100\% QQQ & \pct{9.51} & 0.60 & \pct{-14.04} & \pct{100.00} \\
2015 & 100\% DIA & \pct{0.09} & 0.08 & \pct{-13.84} & \pct{0.00} \\
2015 & 50/50 QQQ-DIA & \pct{4.77} & 0.37 & \pct{-13.55} & \pct{50.00} \\
2016 & RGRR & \pct{13.72} & 0.96 & \pct{-11.06} & \pct{53.30} \\
2016 & 100\% QQQ & \pct{7.22} & 0.51 & \pct{-12.12} & \pct{100.00} \\
2016 & 100\% DIA & \pct{16.39} & 1.28 & \pct{-8.47} & \pct{0.00} \\
2016 & 50/50 QQQ-DIA & \pct{11.82} & 0.88 & \pct{-10.16} & \pct{50.00} \\
2017 & RGRR & \pct{29.03} & 3.48 & \pct{-2.57} & \pct{41.59} \\
2017 & 100\% QQQ & \pct{32.94} & 2.81 & \pct{-4.91} & \pct{100.00} \\
2017 & 100\% DIA & \pct{28.23} & 3.89 & \pct{-3.21} & \pct{0.00} \\
2017 & 50/50 QQQ-DIA & \pct{30.68} & 3.60 & \pct{-2.11} & \pct{50.00} \\
2018 & RGRR & \pct{-4.56} & -0.15 & \pct{-19.43} & \pct{29.92} \\
2018 & 100\% QQQ & \pct{-0.13} & 0.11 & \pct{-22.89} & \pct{100.00} \\
2018 & 100\% DIA & \pct{-3.76} & -0.12 & \pct{-18.15} & \pct{0.00} \\
2018 & 50/50 QQQ-DIA & \pct{-1.81} & 0.01 & \pct{-20.41} & \pct{50.00} \\
2019 & RGRR & \pct{31.41} & 1.98 & \pct{-9.00} & \pct{55.48} \\
2019 & 100\% QQQ & \pct{39.16} & 2.12 & \pct{-11.03} & \pct{100.00} \\
2019 & 100\% DIA & \pct{25.04} & 1.85 & \pct{-6.70} & \pct{0.00} \\
2019 & 50/50 QQQ-DIA & \pct{32.02} & 2.07 & \pct{-8.59} & \pct{50.00} \\
2020 & RGRR & \pct{28.99} & 0.92 & \pct{-29.25} & \pct{45.04} \\
2020 & 100\% QQQ & \pct{48.42} & 1.29 & \pct{-28.65} & \pct{100.00} \\
2020 & 100\% DIA & \pct{9.54} & 0.43 & \pct{-36.71} & \pct{0.00} \\
2020 & 50/50 QQQ-DIA & \pct{28.04} & 0.89 & \pct{-32.28} & \pct{50.00} \\
2021 & RGRR & \pct{24.31} & 1.79 & \pct{-5.16} & \pct{38.15} \\
2021 & 100\% QQQ & \pct{27.42} & 1.42 & \pct{-10.86} & \pct{100.00} \\
2021 & 100\% DIA & \pct{20.83} & 1.60 & \pct{-6.37} & \pct{0.00} \\
2021 & 50/50 QQQ-DIA & \pct{24.42} & 1.67 & \pct{-5.64} & \pct{50.00} \\
2022 & RGRR & \pct{-15.23} & -0.62 & \pct{-23.61} & \pct{30.71} \\
2022 & 100\% QQQ & \pct{-32.68} & -1.07 & \pct{-34.83} & \pct{100.00} \\
2022 & 100\% DIA & \pct{-7.04} & -0.27 & \pct{-20.76} & \pct{0.00} \\
2022 & 50/50 QQQ-DIA & \pct{-20.60} & -0.79 & \pct{-26.80} & \pct{50.00} \\
2023 & RGRR & \pct{37.61} & 2.26 & \pct{-9.20} & \pct{63.19} \\
2023 & 100\% QQQ & \pct{55.40} & 2.56 & \pct{-10.78} & \pct{100.00} \\
2023 & 100\% DIA & \pct{16.16} & 1.37 & \pct{-8.56} & \pct{0.00} \\
2023 & 50/50 QQQ-DIA & \pct{34.64} & 2.27 & \pct{-9.10} & \pct{50.00} \\
2024 & RGRR & \pct{21.40} & 1.42 & \pct{-9.98} & \pct{61.75} \\
2024 & 100\% QQQ & \pct{25.75} & 1.37 & \pct{-13.56} & \pct{100.00} \\
2024 & 100\% DIA & \pct{14.82} & 1.28 & \pct{-6.05} & \pct{0.00} \\
2024 & 50/50 QQQ-DIA & \pct{20.46} & 1.47 & \pct{-8.93} & \pct{50.00} \\
2025 & RGRR & \pct{18.06} & 0.92 & \pct{-20.23} & \pct{60.27} \\
2025 & 100\% QQQ & \pct{20.96} & 0.92 & \pct{-22.77} & \pct{100.00} \\
2025 & 100\% DIA & \pct{14.83} & 0.91 & \pct{-15.83} & \pct{0.00} \\
2025 & 50/50 QQQ-DIA & \pct{18.10} & 0.96 & \pct{-19.15} & \pct{50.00} \\
2026 & RGRR & \pct{29.00} & 1.64 & \pct{-9.66} & \pct{54.46} \\
2026 & 100\% QQQ & \pct{35.52} & 1.52 & \pct{-11.72} & \pct{100.00} \\
2026 & 100\% DIA & \pct{22.67} & 1.55 & \pct{-9.76} & \pct{0.00} \\
2026 & 50/50 QQQ-DIA & \pct{29.39} & 1.68 & \pct{-9.66} & \pct{50.00} \\
\bottomrule
\end{longtable}
\endgroup

\subsection{Output Files}

The replication package is the GitHub repository listed on the first page. It contains the final feature panel, screened signal tables, RGRR rolling outputs, and the reproduction script used to regenerate the reported policy comparison.

\begin{table}[H]
\centering
\caption{Reproducibility Manifest for RGRR Paper Tables}
\label{tab:output-manifest}
\begingroup
\tiny
\setlength{\tabcolsep}{3pt}
\begin{tabular}{p{6.2cm}p{7.2cm}}
\toprule
Artifact & Paper Role\\
\midrule
Feature panel & Daily QQQ/DIA returns, relative states, macro states, and screened score columns used by the RGRR reproduction script.\\
Main-effect screen & Globally screened main effects; defines the single admitted main signal, \code{rate_relief}.\\
Interaction screen & Globally screened interaction set before the incremental third-order filter.\\
Third-order candidate ranking & Candidate ranking by incremental rolling OOS Sharpe versus the main+second-order base.\\
Retained third-order terms & The final two retained third-order interactions used by RGRR.\\
Policy comparison & Final rolling OOS RGRR results and aligned baselines across requested starts.\\
Baseline deltas & Delta table versus 100\% QQQ, 100\% DIA, and 50/50 QQQ--DIA.\\
OOS selections & Rolling block-level lambda selection and selection-stability evidence.\\
Daily path & Daily RGRR net return, turnover, QQQ weight, and selected configuration identifier.\\
Generated report & Stand-alone Markdown report generated from the reproduction script.\\
\bottomrule
\end{tabular}
\endgroup
\end{table}

The method-name mapping is:
\begin{quote}
RGRR in the paper = \code{filtered_ix3_top2} in the CSV/report artifacts.
\end{quote}
This mapping is included only for reproducibility. The paper avoids using implementation labels as method names because they describe code mechanics rather than the economic idea.

\section{Start-Date Sensitivity Selection Stability}

The cash-overlay study includes a start-date sensitivity section that separates two questions: whether the final method works across requested starts, and whether the walk-forward selector is simply choosing one static configuration \cite{xiong2026cashoverlay}. The same audit is included here. Table~\ref{tab:appendix-selection-stability} reports the number of 63-day OOS blocks, the number of distinct archived lambda configurations selected, and the share of the most frequent configuration. The configuration identifiers are archive labels only; they are not method names.

\begin{table}[H]
\centering
\caption{Walk-Forward Selection Stability for RGRR}
\label{tab:appendix-selection-stability}
\begingroup
\tiny
\setlength{\tabcolsep}{3pt}
\resizebox{\textwidth}{!}{%
\begin{tabular}{lrrlrrrr}
\toprule
Period & Blocks & Unique Configs & Top Archive Config & Top Blocks & Top Share & Median Train Sharpe & Median Train Turnover\\
\midrule
2008 start & 68 & 17 & \code{filtered_ix3_top2_015} & 17 & \pct{25.00} & 0.91 & \pct{531.19}\\
2018 OOS & 32 & 12 & \code{filtered_ix3_top2_035} & 8 & \pct{25.00} & 0.97 & \pct{466.61}\\
2020 OOS & 26 & 8 & \code{filtered_ix3_top2_035} & 8 & \pct{30.77} & 0.83 & \pct{513.34}\\
2022 OOS & 18 & 6 & \code{filtered_ix3_top2_007} & 5 & \pct{27.78} & 0.80 & \pct{452.06}\\
\bottomrule
\end{tabular}
}
\endgroup
\end{table}

The top selected configuration never dominates the majority of blocks. This is expected because the rolling selector is allowed to adapt the lambdas on the fixed RGRR signal groups. The important restriction is that the economic vocabulary does not change: the main effect, second-order interactions, retained third-order interactions, mapping parameters, cost convention, and turnover penalty are all fixed before the rolling OOS path is evaluated.

This stability audit also clarifies the distinction between rolling OOS and a fixed-parameter holdout. A fixed-parameter holdout would select one set of lambdas before the OOS window and hold it unchanged. RGRR instead uses a walk-forward design: every 63-day test block is preceded by a 756-day training window, and only lambdas are re-selected. The method therefore tests whether a fixed, statistically screened signal universe can remain useful while its group weights adapt to recent evidence.

Finally, the stability results explain why the paper reports turnover so prominently. Median training turnover is above the 300\% penalty threshold in every requested start, even after adding the penalty. This does not invalidate the signal result, but it constrains the implementation claim. The current version demonstrates a statistically disciplined and risk-adjusted relative-rotation mechanism; a production version would need additional trading gates or slower execution rules to reduce realized turnover.

\end{document}